\documentclass[format=acmsmall, review=false, screen=true]{acmart}

\usepackage[para]{footmisc}

\usepackage{amsmath}
\usepackage{natbib}
\usepackage{amsfonts}
\usepackage{graphicx}
\usepackage{hhline}
\usepackage{hyperref}
\usepackage{enumerate}
\usepackage{multirow}
\usepackage{array}
\usepackage[para]{footmisc}
\usepackage{adjustbox}
\usepackage{balance}
\usepackage{subcaption}
\usepackage[normalem]{ulem}
\usepackage[linesnumbered,boxed,ruled,commentsnumbered]{algorithm2e}
\usepackage[show]{chato-notes}
\usepackage{marginnote}

\newcommand{\xiw}[1]{\textcolor{black}{#1}}
\newcommand{\ziy}[1]{\textcolor{black}{#1}}
\newcommand{\io}[1]{\textcolor{black}{#1}}
\newcommand{\cm}[1]{\textcolor{black}{#1}}
\newcommand{\zixuan}[1]{\textcolor{black}{#1}}
\newcommand{\zx}[1]{\textcolor{black}{#1}}
\newcommand{\zxy}[1]{\textcolor{black}{#1}}
\newcommand{\zxx}[1]{\textcolor{black}{#1}}
\newcommand{\yo}[1]{\textcolor{black}{#1}}
\newcommand{\yzx}[1]{\textcolor{black}{#1}}
\newcommand{\yy}[1]{\textcolor{black}{#1}}
\newcommand{\yz}[1]{\textcolor{black}{#1}}
\newcommand{\final}[1]{\textcolor{black}{#1}}

\AtBeginDocument{%
  \providecommand\BibTeX{{%
    \normalfont B\kern-0.5em{\scshape i\kern-0.25em b}\kern-0.8em\TeX}}}

\setcopyright{acmlicensed}
\acmJournal{TOIS}
\acmYear{2023} \acmVolume{1} \acmNumber{1} \acmArticle{1} \acmMonth{1} \acmPrice{15.00}\acmDOI{10.1145/3618298}

\begin{document}

\title{Contrastive Graph Prompt-tuning for Cross-domain Recommendation}

\author{Zixuan Yi}
\email{z.yi.1@research.gla.ac.uk}
\affiliation{%
 \institution{University of Glasgow}
 \city{Glasgow}
 \state{Scotland}
 \country{United Kingdom}
}


\author{Iadh Ounis}
\email{iadh.ounis@glasgow.ac.uk}
\affiliation{%
 \institution{University of Glasgow}
 \city{Glasgow}
 \state{Scotland}
 \country{United Kingdom}
}

\author{Craig Macdonald}
\email{craig.macdonald@glasgow.ac.uk}
\affiliation{%
 \institution{University of Glasgow}
 \city{Glasgow}
 \state{Scotland}
 \country{United Kingdom}
}

\fancyhead{}
\begin{abstract}
Recommender systems commonly suffer from the long-standing data sparsity problem \final{where insufficient user-item interaction data limits the systems' ability to make accurate recommendations.}
This problem can be alleviated using cross-domain recommendation techniques.
In particular, in a cross-domain setting, knowledge sharing between domains \cm{permits improved effectiveness on the target domain. While recent cross-domain \final{recommendation} techniques used a pre-training configuration,} we argue that such techniques lead to a low fine-tuning efficiency, especially when using large neural models. 
\cm{In recent language models, {\em prompts} have been used for \final{parameter-efficient and time-efficient tuning} of the models on \yz{the} downstream tasks - these prompts represent a tunable latent vector \io{that} permits \yz{to freeze} the rest of the language model's parameters.}
To address the cross-domain recommendation task in an efficient manner, we propose a novel \yz{Personalised} Graph Prompt-based Recommendation (PGPRec) framework, which leverages the efficiency \final{benefits} \yz{from} prompt-tuning. 
In such a framework, we develop personalised and item-wise graph \io{prompts} based on \io{relevant} items to those \yz{items the user has} interacted \io{with}.
\zx{In particular, we apply Contrastive Learning (CL) to generate the pre-trained embeddings, \final{to allow} an increased generalisability in \io{the} pre-training stage and \io{to} ensure \final{an} effective \final{prompt-tuning} stage.}
\ziy{To evaluate the effectiveness of our PGPRec framework in a  cross-domain setting, we conduct \io{an} extensive evaluation \io{with} 
the top-$k$ recommendation \io{task} and perform a cold-start analysis. The \io{obtained} empirical results on four Amazon Review datasets show that our proposed PGPRec \io{framework} can reduce up to 74\% of \io{the} \zxx{tuned} parameters with a competitive performance and \yz{achieves an} 11.41\% improved performance compared to the strongest baseline in a cold-start scenario.}

\end{abstract}




\maketitle

\section{Introduction}\label{sec:intro}
Personalised recommendation techniques, which learn user preferences and find items \zixuan{related to} their interest, have been \zixuan{increasingly} developed in the last decades.
In particular, for a recommender, the effective \zixuan{personalisation of the} recommendation results, frequently rely on rich available data, such as historical user-item interactions, domain knowledge, as well as \zixuan{the} user demographics and profiles~\cite{kg2014survey}.
However, in the recommendation literature, the performance of a personalised recommender system \zixuan{deployed on} a single domain, \final{often} suffers from the commonly \zixuan{observed} data sparsity issue. 
\final{This refers to an insufficient number of user-item interactions, which hinders accurate recommendation generation~\cite{man2017cross}.}
Therefore, Cross-Domain Recommendation (CDR)~\cite{tang2016cross,zang2021survey}, which relies \zixuan{on} shared users \zixuan{over} pairs of domains to transfer relevant information from the source domain to the target domain, is \zixuan{typically} used to alleviate the negative effect of sparse interactions, so as to improve the recommendation performance when applied to a \zixuan{different} target domain.

Inspired by recent advances in natural language
processing~\cite{devlin2018bert} \zixuan{centred on the paradigm of first} pre-training then fine-tuning,
\zx{some} cross-domain recommendation techniques~\cite{wang2021pre, zhu2022personalized} \zixuan{considered} the application of \yz{this} pre-training \yz{and fine-tuning} mechanism. 
\yzx{In particular, a common development routine for shared users in cross-domain recommendation (CDR) involves first \yz{the} pre-training \yz{of} the recommendation model using data from the source domain to learn domain-invariant knowledge \final{(e.g. user profiles)}. Subsequently, the knowledge \final{contained in} the pre-trained model is used to initialise the user/item embeddings in the target domain.}
Such a strategy has been shown to be effective in alleviating the \yz{mentioned} data sparsity issue~\cite{xiao2021uprec}.
Moreover, a myriad of pre-training recommendation models~\cite{wang2021pre,meng2021graph} have been proposed to leverage the structural data from the source domain, \zixuan{which can leverage collaborative filtering signals to yield better user/item representations \final{by leveraging the high-order connectivity via graph convolutional operations\cite{kipf2016semi}}.}
However, the \final{effectiveness of recommendation} on the target domain \zixuan{can} be \zixuan{negatively impacted} \zixuan{if the pre-trained model learns from \zixuan{non-relevant} domain-specific features in the source domain.}
Moreover, \cm{the fine-tuning of a pre-trained model causes all parameters of the model to be updated, even \yz{in the case where} not all parameters need to be updated \io{to obtain} an effective model. This \io{makes} fine-tuning inefficient.}
Therefore, \io{such an} efficiency limitation \io{provides a motivation for devising} a \yz{new} framework \zixuan{that can efficiently transfer effective pre-trained embeddings.} 
\begin{figure}
    \centering
    \small
    \includegraphics[trim={2cm 6cm 0cm 6cm},clip,width=16cm]{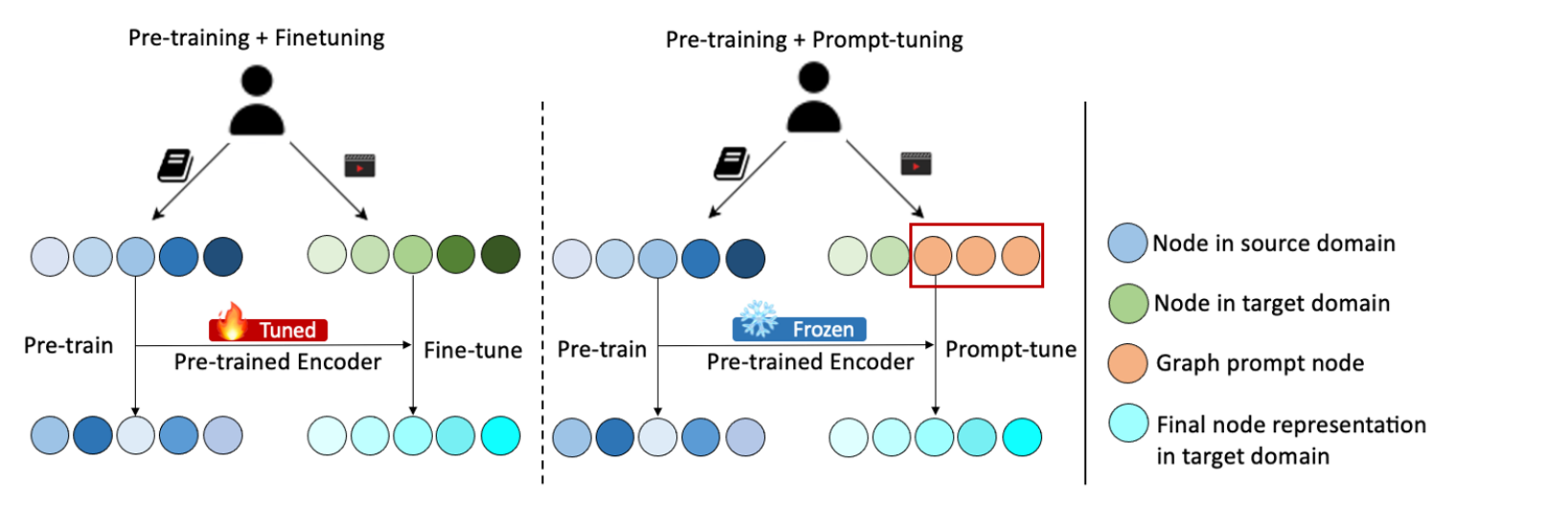}
    \caption{\ziy{Overview of conventional graph fine-tuning and personalised graph prompt-tuning in \io{Cross-Domain Recommendation  (CDR)}.}}  
 \label{fig:apr-cdr-solution}
\end{figure}

Recently, in the Natural Language Processing (NLP) field, prompt-tuning~\cite{brown2020language, lester2021power} has been widely applied \zixuan{to address} many NLP tasks and has shown its usefulness in extracting useful information \cm{from a pre-trained model, such that it can be adapted to a downstream task with less \io{efforts}.}
\zixuan{Several} prompt-variants \zixuan{have been proposed}, such as term filling-based text templates~\cite{brown2020language,gao2020making} or continuous embeddings (i.e.\ soft prompt)~\cite{li2021prefix, qin2021learning}, 
\zixuan{which \yo{effectively} bridge the pre-training \yz{process in various natural language processing tasks}}.
\yzx{In adapting prompt-tuning for recommender systems, 
several studies~\cite{wu2022personalized,cui2022m6,geng2022recommendation} have investigated the application of prompt techniques within the recommendation context,
focusing on leveraging user and item information to generate customised prompts that can effectively address specific recommendation tasks.}
For example, Wu et al.~\cite{wu2022personalized} proposed \zixuan{a} soft prefix prompt based on \final{users'} profiles to tackle the cold-start problem in a sequential recommendation setting. Another \zixuan{stream of work}~\cite{cui2022m6,geng2022recommendation} attempted to convert the available data resources, such as the user-item interactions, \io{the} user descriptions, \io{the} item metadata, and \io{the} user reviews, into natural language sequences. \zixuan{Next}, they used a composition of such sequences as \yz{a} prompt and the \io{pre-trained} models for efficiently 
addressing various downstream tasks of the recommender system \zixuan{\zixuan{(e.g., cold-start recommendation, few/zero-shot recommendation and user profile prediction)}.}
\yzx{The aforementioned examples \yo{show} that existing prompt tuning-based recommendation techniques rely on sequential patterns \yz{in the users'} interacted items and mimic NLP models designed for sequential text.}
\yy{This demonstrates the effective application of \yo{the} prompt-tuning techniques in the context of recommendation systems.}

\zixuan{Inspired by the success of prompt-tuning in sequential recommendation, we  introduce the prompt-tuning paradigm into Graph Neural Networks (GNN)-based recommendation models to alleviate \zixuan{their} low efficiency \zixuan{limitation}, \final{as highlighted in}~\cite{liu2021eignn}.}
However, \zixuan{it is challenging to design a prompt-tuning technique \zixuan{that} makes \zixuan{a} full use of knowledge from structural data rather than \zixuan{from} sequential data in the existing recommendation techniques.}
Therefore, it is essential to bridge the GNN recommendation models with the existing prompt-tuning techniques, \yz{which use} a sequential modelling of data. 
In this \zixuan{paper}, we show that the language models in the NLP domain could be a special form of \zixuan{a} GNN (\io{as will be} detailed in Section~\ref{sec:eqv}), which allows us to propose novel templates \zixuan{for} the personalised graph \yz{prompts}. \yz{These templates} \io{could then} be applied to GNN-based recommenders.
\final{As an illustrative example, Figure 1 shows how our proposed prompt-tuning framework contrasts with conventional fine-tuning approaches in a CDR setting.}
\io{The} existing CDR solutions \io{typically} pre-train a graph encoder in a source domain. Then, they fine-tune the well-trained graph encoder with the interaction data for recommendation in \io{the target} domain. \io{Therefore, in this paper, we} propose a prompt-tuning framework instead of \io{the} fine-tuning \yz{process}, to improve the tuning efficiency by leveraging personalised graph prompts while \io{fixing} the pre-trained \io{parameters}.
To be more specific, we build personalised graph prompts with a series of \zixuan{relevant} items, \zixuan{which \zixuan{we call the neighbouring-items}\footnote{
Given an item, we denote by its "neighbouring-items" those items that share its same attributes in the target domain, i.e. the set of items with the same attributes.
In this work, we define the \zixuan{neighbouring-items} according to three types of available metadata -- i.e.\ also\_bought, also\_viewed and bought\_together -- and refer \zixuan{to} them \zixuan{as} `neighbouring-items' throughout this paper.} and leverage these prompts to efficiently \yz{enhance} the final user/item representations.}
\zixuan{Hence}, we argue that, in a CDR setting, GNN recommenders \zixuan{enriched} with our proposed graph prompt can efficiently \yz{leverage} the learned knowledge from the source domain \zixuan{to} effectively \yz{improve} the recommendation \zixuan{performance} in the target domain.
Furthermore, we propose a novel Personalised Graph Prompt-based Recommendation (PGPRec) framework, which encapsulates our proposed \yz{personalised prompts} as well as a \zixuan{Contrastive Learning (CL)-empowered} GNN recommender for cross-domain recommendation. \zixuan{Indeed, with the integration of a graph-based contrastive learning, the pre-trained model \zixuan{enforces} the divergence of the learned user/item embeddings~\cite{wu2021self}, which makes them generalisable to various target domains\footnote{In this paper, we also evaluate the contribution of using the contrastive learning \zx{through the application of} PGPRec to distinct source-target domain pairs.} with \zixuan{an} improved recommendation performance.}

To summarise, our contributions are \final{four-fold}: 
\begin{itemize}
    \item We propose a personalised graph prompt-based recommendation framework for cross-domain recommendation, 
    \zixuan{which uses contrastive learning to enhance the user/item representations in both \zixuan{the} pre-training and tuning stages}
    and improves the efficiency of the tuning phase.
    \item We introduce item-wise personalised prompts to be used in a user-item bipartite graph to effectively \yz{enhance} the user/item representations in the target domain.
    \item We conduct extensive experiments on four Amazon Review datasets. \yz{We} perform a TOST test~\cite{schuirmann1987comparison} to demonstrate that PGPRec shows a significantly improved efficiency in comparison to the state-of-the-art GNN recommenders \yz{while achieving a comparable recommendation performance.}
    \item \ziy{We further conduct a \io{detailed} analysis \io{of the} cold-start users in the target domain \io{using} four Amazon Review datasets. \yz{Our findings show that PGPRec significantly outperforms the strongest baseline in a cold-start scenario.}}

\end{itemize}
 
The remainder of this paper is organised as follows: In Section~2, we position our proposed \yz{PGPRec} framework in the literature. Section~3 \zixuan{describes} \yz{our} methodology \yz{and the} specific implementations of \yz{the} personalised graph prompts during \zixuan{the} training stage as well as the optimisation objective of the cross-domain recommendation task. The experimental setup and the results of our empirical experiments are presented in Section~4, followed by concluding remarks and future work in Section~5.

\section{Related Work}\label{sec:rwork}
In this section, we discuss related methods and techniques to \zixuan{our conducted study}, \zixuan{namely} pre-training recommenders, cross-domain recommenders and prompt-tuning techniques.
\subsection{Pre-training in Recommendation}
Recently, pre-training techniques, which \zixuan{learn} knowledge from large-scale datasets for an improved model performance, have achieved \zixuan{several} successes in addressing the recommendation task~\cite{zeng2021knowledge,yi2022multi}. 
\yzx{A typical approach for learning a recommendation model \yo{using the pre-training techniques} involves \yo{first the initialisation of} the model with knowledge obtained from a pre-training stage and subsequently \yo{the} fine-tuning \yo{of the model} using supervised signals from the target recommendation scenario.
This pre-training and fine-tuning paradigm enables the model to effectively incorporate prior knowledge while adapting to the specific characteristics of the target domain~\cite{zeng2021knowledge}.}
For example, BERT4Rec~\cite{sun2019bert4rec} pre-trains \final{a} BERT \final{transformer architecture} to learn \zixuan{a} sequential pattern of items to model \io{the} user \zxx{behaviour} sequences, \io{thereby leading} \zixuan{to a} promising performance in sequential recommendation.
Similarly, 
ASReP~\cite{liu2021augmenting} pre-trains a transformer 
to generate fabricated historical items at the beginning of short sequences and then fine-tune\final{s} the transformer based on the new sequences to alleviate the cold-start problem. 

\ziy{\io{On the other hand}, there \io{have been} several efforts to pre-train recommenders by designing self-supervised \io{auxiliary} tasks to discover \io{supervised} signals from the raw data. Contrastive learning recommendation is \io{increasingly considered to be} \io{a} promising \io{approach within the family of} self-supervised learning recommendation \io{approaches}~\cite{yu2022self,yi2023graph}, which can be applied for various recommenders by perturbing the raw data. \io{For example,}
CL4SRec~\cite{xie2022contrastive} performs masking, cropping and reordering \io{operations} to augment the input \io{items} sequence then \io{contrasts} the augmentations to pre-train a transformer-based encoder for sequential recommendation. Another example of \io{contrastive learning recommendation model is} PCRec~\cite{wang2021pre}, \io{which} also uses contrastive learning to enhance the cross-domain recommendation. \io{PCRec} pre-trains a GIN encoder~\cite{xupowerful} by contrasting \io{a} random walk-based augmentation in the source domain and then transfers the pre-trained user/item \io{embeddings} to \yy{initialise} a MF model in \io{the} target domain. As \io{illustrated \final{by} the aforementioned models}, contrastive learning has demonstrated its ability in learning generalisable representations in pre-training recommendations.}

Another line of work~\cite{xiao2021uprec,yuan2020parameter,meng2021graph} attempted to \ziy{use extra knowledge }to enhance the positive effect of pre-training.
For example, UPRec~\cite{xiao2021uprec} encapsulates various pre-training tasks based on user attributes and social relations to learn comprehensive user representations for an improved user-item sequence modelling.
Another example, PeterRec~\cite{yuan2020parameter}, also pre-trains the user representations based on \io{the} user-item interactions, but applies the learned model to a different domain (i.e. the target domain in a cross-domain setting).
As \zixuan{\final{can} be} seen \final{from} \io{the} above work, cross-domain recommendation is well-aligned with the pre-training and fine-tuning \zixuan{paradigm}, which aims to transfer useful knowledge from the source domain to benefit the effectiveness of \zixuan{the} resulting recommendations in the target domain.
Moreover, the existing \zixuan{Cross-Domain \io{Recommendation} (CDR)} models \zixuan{typically} transfer knowledge learned from sequential patterns \zixuan{while} very few approaches attempt to benefit from the learned structural \zixuan{knowledge~\cite{qiu2020gcc}.}
Therefore, \zixuan{\zixuan{differently} from \io{the} existing approaches \zixuan{that learn} from sequential data, we propose PGPRec to leverage \ziy{contrastive learning to pre-train the GNN-based recommenders.} \zixuan{Our proposed PGPRec framework extracts} intrinsic and structural knowledge from the pre-trained models in a cross-domain recommendation setting.}

\subsection{Cross-domain Recommendation}\label{sec:cdr}
\final{In general, cross-domain recommendation (CDR) is an application of transfer learning to recommendation scenarios involving two domains, namely a {\em source} domain (which provides us with additional useful knowledge, in order to reduce data sparsity and therefore improve recommendations on a separate {\em target} domain. While different CDR scenarios exist, namely shared-users, shared-items and shared-nothing},\footnote{Actually, improvements identified in shared-nothing cross-domain recommendation have  been shown to simply be due to increased model capacity~\cite{manotumruksa2019cross}.} \final{in this paper we focus upon shared-users, where user profiles on the source domain are used to permit improved personalisation of recommendations in the target.}
For example, CMF~\cite{tang2016cross} -- a classical CDR approach -- jointly factorises the rating matrices from two domains with a shared global user embedding matrix. 
Another \io{approach}, CoNet~\cite{hu2018conet}, introduces a cross-connection unit to transfer the user-item interaction features between two domains.
\ziy{Similarly, CBMF~\cite{mirbakhsh2015improving} uses a cluster-based matrix to learn the correlation between \io{the} user clusters and \io{the} item clusters in different domains and then uses this matrix to alleviate the cold-start problem.}
With the rise of Graph Neural Networks (GNNs) in recommender systems, PPGN~\cite{zhao2019cross} adopts \zixuan{the GNN technique} to explore the high-order connectivity between users and items on a joint interaction graph of two domains \zixuan{so as to} allow the knowledge-transfer with shared user features.

Among the CDR approaches, \zixuan{those} in the literature~\cite{wang2021pre,yuan2019darec,zhu2022personalized,zeng2021knowledge} that \zixuan{have} focused on pre-training in a source domain and fine-tuning in the target domain \zixuan{are the most relevant} to our proposed PGPRec framework.
For instance, 
\ziy{EMCDR~\cite{man2017cross} pre-trains a multi-layer \io{perceptron} (MLP) as a mapping function in a source domain and then \io{transfers} the learned user representation into the target domain for cold-start users.}
Similarly, PCRec~\cite{wang2021pre} first pre-trains a graph encoder and then transfers the learned user features to the target domain for an improved performance after fine-tuning the resulting recommendation model.
However, \zixuan{the} existing approaches assume a prediction objective during pre-training on \zixuan{the} source domains, which could lead to sub-optimal representations \zixuan{that can} impede the effectiveness of the resulting recommender models~\cite{wang2021pre}. 
Unlike \zixuan{these} existing approaches, we adopt contrastive pre-training to encourage the pre-trained models to learn more \io{generalisable} features,
\yzx{with an emphasis on transferring domain-invariant knowledge.
}
\yzx{This \yy{contrastive pre-training strategy allows for} a higher divergence in the representation of learned features from the source domain, providing the model with a better initialisation point for diverse target domains. By focusing on the transfer of domain-invariant knowledge, our approach fosters the development of recommendation systems that can effectively harness the power of cross-domain techniques, thereby contributing to the field of domain adaptation in recommendation systems.}
\yzx{Beyond addressing the issue of sub-optimal training objectives,
the pre-training and fine-tuning paradigm 
\yo{has} a critical limitation when applied to cross-domain recommendation, \yo{namely} its low efficiency~\cite{chen2022revisiting}. \yo{Indeed, the} necessity to train the model twice results in significant time and computational costs associated with \yo{the model's} parameters.}
\yy{To address this low-efficiency challenge,
we \yo{propose} the PGPRec framework, a tailored solution for cross-domain recommendation systems, \yo{which} uses a personalised graph prompt-tuning technique to adapt \yo{the} users' \yo{preferences} in the target domain.}
\yzx{By leveraging pre-trained knowledge and personalised graph prompts, PGPRec provides a more efficient solution for cross-domain recommendation.}
\ziy{ 
\yy{On the other hand,}
\yy{in order} to 
investigate the cold-start problem in cross-domain recommendation, we also leverage contrastive pre-training and personalised graph \yo{prompts} to benefit the cold-start users, \io{i.e. those users that} have sparse interactions in a target domain.}

\subsection{Prompt-tuning}\label{sec:pt}

\io{A prompt} originally refers to \io{a} prefixed plain text and is combined with the input of the pre-trained models for an improved semantic understanding of the task~\cite{radford2019language}. For example, GPT-3~\cite{brown2020language} leverages manually devised prompts for transfer learning in \zixuan{Natural Language Processing (NLP)}. 
With the follow-up work on prompt-variants~\cite{shin2020autoprompt,jiang2020can}, there \io{has been} a growing number of recent prompt design methods, including hard and soft prompts, following the ``\textit{pre-train, prompt, and predict}" paradigm~\cite{liu2021pre}. Hard prompts are discrete textual terms~\cite{schick2020exploiting,brown2020language} \zixuan{while} soft prompts are one or many continuous \yz{learned} embeddings~\cite{li2021prefix,lester2021power,liu2021p}. In particular, soft prompts are randomly initialised and then optimised via parameter-tuning.
Consequently, prompt-tuning, using soft prompts, narrows the gap 
between the objectives of the source tasks and \zixuan{that of} the downstream tasks by transforming the inputs to the target model in a certain format~\cite{wu2022adversarial}. \io{In addition}, \io{the} prompt-tuning methods only rely on fine-tuning \io{a} small set of parameters within the soft prompts to achieve \io{a} competitive performance compared to \zxx{the} fine-tuned models, \zixuan{which endows the final user/item representations in an efficient manner.}

To the best of our knowledge, only a limited number of work \io{have recently} applied prompt-tuning to recommender systems. 
\ziy{In order to address various downstream tasks in recommender systems,} 
\cite{cui2022m6,geng2022recommendation} attempted to follow the techniques from the NLP domain in leveraging the prompt-tuning paradigm. M6-Rec~\cite{cui2022m6} \final{considered} the user behaviour as a sequence of `text' and \final{addressed} both the click-through-rate (CTR) prediction and the explainable recommendation \io{tasks} 
with the assistance of randomly initialised soft prompts~\cite{geng2022recommendation}. 
Similarly, P5~\cite{geng2022recommendation} 
convert\final{ed} \io{the} user descriptions, \io{the} item metadata, and \io{the} user reviews to natural language sequences as prompts to effectively leverage a pre-trained transformer model. 
On the other hand, Wu et al.~\cite{wu2022personalized} treated cold-start recommendation as the downstream task of sequential recommendation and enhanced the performance by building prefixed soft prompts based on \io{the} user profiles. 
Although prompt-tuning has also been applied to sequential recommendations, 
\zixuan{the} existing approaches relied on the sequential pattern of the input data.
In this work, \zixuan{we argue that another possible} strategy is to transfer the structural pattern of a user from \zixuan{the} user-item bipartite graphs.
Indeed, we introduce \zixuan{novel} personalised graph prompt\zx{s}, which leverage the
\zx{neighbouring-items and the tunable continuous vectors as hard and soft graph prompts},
to address cross-domain recommendation with improved \final{training} efficiency.
\ziy{Moreover, to investigate the characteristics of prompt-tuning on \io{graphs}, we \io{ensure that} PGPRec \io{leverages} the pre-trained knowledge and informative personalised graph prompts to benefit the cold-start users, which have sparse interactions in a target domain.}



\section{Methodology}
In this section, we first present \zixuan{the} cross-domain recommendation task (Section~\ref{PDN}). Next, we describe
the composition of \zixuan{our proposed} personalised graph prompt\zx{s} and training process (Section~\ref{sec:frame}). 
We also \zixuan{describe} how to derive the personalised graph prompt-tuning paradigm by \zixuan{justifying} the equivalence between \zixuan{the} transformer and GNN (Section~\ref{cpgp}), along with \zixuan{the} detailed implementations and objectives during \zixuan{the} training phases (Section~\ref{sec:cl4rec}).


\subsection{Preliminaries} \label{PDN}
In this paper, we focus on addressing the ranking-based recommendation task in a cross-domain setting. 
Conceptually, we consider two domains, the source domain $D_{\mathrm{S}}$ and the target domain $D_{\mathrm{T}}$. The set of users in both domains is shared, \io{and} denoted by $U$ (of size $m=|U|$). 
Let the set of items in 
$I_{\mathrm{S}}$ (of size $n_{s}=\left|D_{\mathrm{S}}\right|$)
and $I_{\mathrm{T}}$ (of size $n_{t}=\left|D_{\mathrm{T}}\right|$), respectively. 
Then, we aim to make effective and efficient recommendations to users \io{in} $U$ with a ranked list of items from the target domain $D_{\mathrm{T}}$
, \yz{while} leveraging the learned knowledge from the source domain $D_{\mathrm{S}}$.
In particular, \zixuan{we} devise two bipartite graphs \final{$\mathcal{G}_{S}$ and $\mathcal{G}_{T}$} for both \zixuan{the} source and target \zixuan{domains},
where \zixuan{the} nodes represent users/items and \zixuan{the} edges indicate interactions between \io{the} users and items. Formally speaking, with \io{the} interaction graphs $\mathcal{G}_{S}$ and $\mathcal{G}_{T}$, we pre-train a graph encoder $f$ in a selected source domain $D_{\mathrm{S}}$ and adapt it \zixuan{to} the target domain $D_{\mathrm{T}}$ for an enhanced knowledge of \io{the} user preferences \zixuan{so as} to effectively and efficiently recommend the top-$k$ items related to their interests.
\ziy{The notations that we will use throughout this paper are \io{summarised} in Table~\ref{tab:nota}.}

\begin{table}[!h] 
\begin{center}
\caption{\ziy{Notations \io{used in this paper to describe the proposed PGRec framework}.} }
\begin{tabular}{c l}
\cline{1-2}
\hline
Symbol & Description\\
\hline
$D_{\mathrm{S}}$ & the source domain  \\ 

$D_{\mathrm{T}}$ & the target domain  \\ 

$I_{\mathrm{S}}$ & the set of items in $D_{\mathrm{S}}$  \\ 

$I_{\mathrm{T}}$ & the set of items in $D_{\mathrm{T}}$  \\ 

$U$ & the set of users which is shared by both \io{the} source domain and \io{the} target domain\\ 

$\mathcal{G}_{S}$ & the user-item interaction graph in $D_{\mathrm{S}}$ \\ 

$\mathcal{G}_{T}$ & the user-item interaction graph in $D_{\mathrm{T}}$ \\ 

$\mathcal{V}$ & the node set \\ 

$\mathcal{E}$ & the edge set \\ 

$M$, $M'$ &  the masking vectors on \io{the} edge set $\mathcal{E}$ \\ 

$N_{u}$ & the neighbour nodes of user $u$ in the interaction graph\\

$N_{i}$ & the neighbour nodes of item $i$ in the interaction graph\\

$R(u)$ & the ground-truth set of items that user $u$ has interacted with\\

$\hat{R}(u)$ & the ranked list of items generated by a recommender\\

$h^{\ell}$ & the node representation at \io{the} $\ell$-th layer  \\ 

$\boldsymbol{e_{u}}$,$\boldsymbol{e_{i}}$ & the embedding vector of the $\ell$-th layer for users and items\\ 

$\boldsymbol{e_{p_h}}$ & the embedding vector of \io{the} neighbouring-item as \io{a} hard prompt\\ 

$\boldsymbol{e_{p_s}}$ & the random initialised embedding vector  as \io{a} soft prompt\\ 

$B$ & the batch size\\ 

$\lambda_{1},\lambda_{2}$ & the regularization term\\ 

$\Theta$ & the parameters of the model \\  

\hline
\end{tabular}
\label{tab:nota}
\end{center}
\end{table}

\subsection{\zixuan{The PGPRec} Framework}\label{sec:frame}
\begin{figure}
    \centering
    \small
    \includegraphics[trim={1cm 4cm 0cm 4cm},clip,width=16cm]{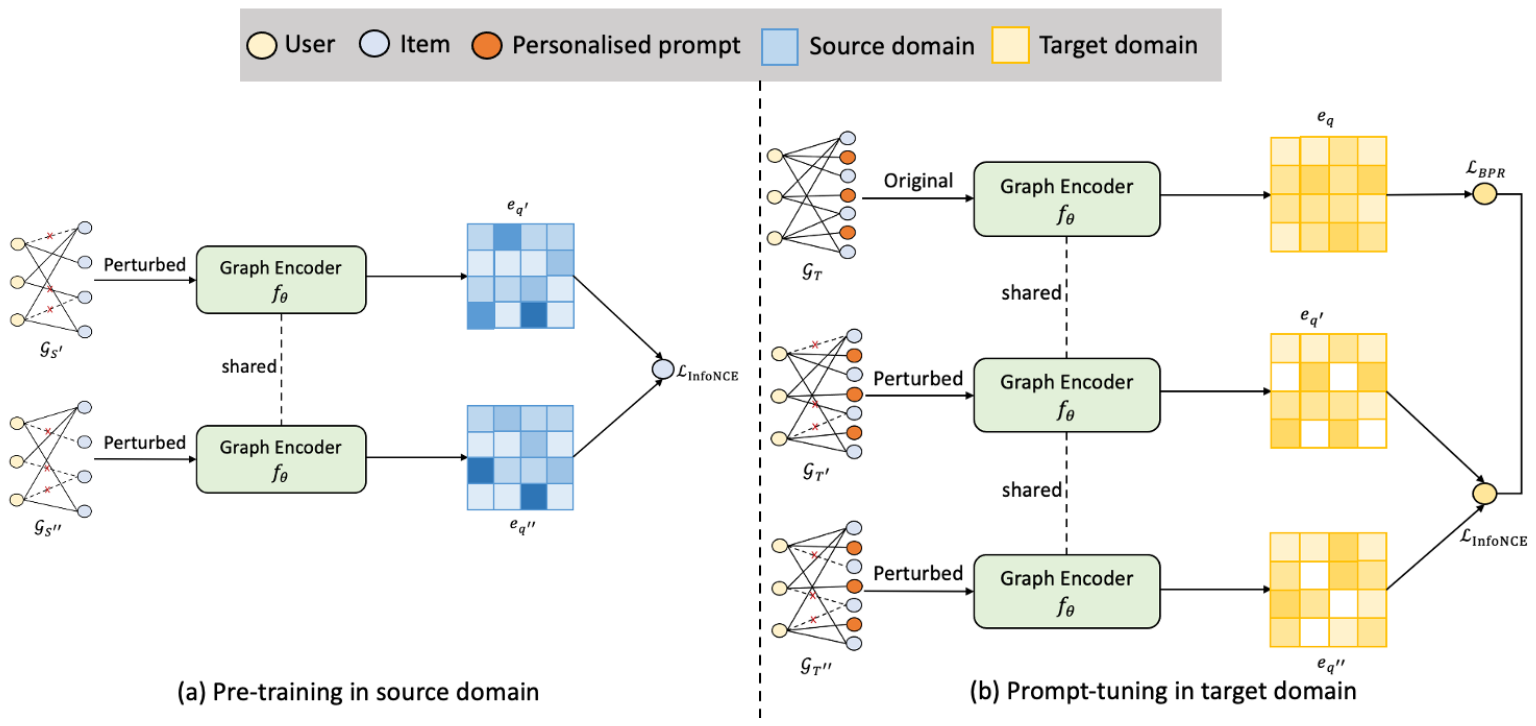}
    \caption{\yzx{An illustration of {the} PGPRec framework {using} a specific example. The framework is mainly composed of two steps. (a) In the pre-training stage, PGPRec trains a GNN model to learn the source domain’s transferable knowledge through a contrastive objective. (b) Then {PGPRec} uses the pre-trained GNN model and {the} personalised graph prompts to enable the prompt-tuning in the target domain through multi-task learning.}}  
 \label{fig:pt}
 \vspace{-4mm}
\end{figure}

First, we provide an overview of our proposed recommendation framework, Personalised Graph Prompt-based Recommendation (PGPRec). 
In particular, Figure~\ref{fig:pt} \zixuan{shows and illustrates} the major components included in \zixuan{our proposed} framework.
Specifically, for our cross-domain
recommender, we adopt \zixuan{the} GNN \zixuan{technique} to model the data from each domain, for capturing the high-order features and allowing the transfer of additional structural knowledge aside from the typical user/item features (e.g.\ \zixuan{the} user profiles and item attributes). 
To instantiate the discussed GNN \zixuan{technique}, we are not \zixuan{limited} to the commonly used GCN model~\cite{kipf2016semi}. \zixuan{Indeed,} PGPRec is also flexible \io{in allowing} to use other GNN-based techniques, such as GAT~\cite{velivckovic2017graph}, NGCF~\cite{wang2019neural} or LightGCN~\cite{he2020lightgcn}.
Inspired by the recent success of applying prompt-tuning in the NLP domain, we also introduce \zx{the} graph prompt\zx{s} to assist the cross-domain recommendation. In particular, \zixuan{as a} rationale \zixuan{for} applying prompt-tuning to \zixuan{a} graph model, we argue that the GNN \zixuan{technique} can be considered as another type of a transformer model. Note that we \io{explain and} justify this argument in Section~\ref{cpgp}. In our PGPRec framework, we propose a mixture of hard and soft graph \io{prompts} \ziy{for personalised recommendation}, \zx{\io{which we collectively denote as} \textit{personalised graph prompts}}. First, we describe the hard graph \zixuan{prompts}. 
For each user, we develop personalised prompt\zx{s} according to the \zixuan{associated} \zixuan{relevant items (denoted as \io{the}} neighbouring-items). 
\zixuan{Given an item, we denote by its "neighbouring-items" those items that share its same attributes in the target domain, i.e. the set of items with the same attributes.}
\zixuan{For example}, in this paper, we refer to \zixuan{the} use of relevant items according to 
\zixuan{three available attributes in \zixuan{the} metadata} (i.e.\ also\_bought, also\_viewed and bought\_together) in the used Amazon datasets. 
\zixuan{However, such} \zixuan{neighbouring-items are} not necessarily fixed and can be \zixuan{naturally} changed \zixuan{or extended} \zixuan{depending on the used} datasets. Next, we consider such \zixuan{neighbouring-items} as \yz{adjacent} nodes to a target user node. For example, \zixuan{consider} a user $u_{0}$ \zixuan{that has} interacted with items $i_{0}$ and $i_{1}$. Other users that interacted with item $i_{0}$ \textit{also\_bought} items $i_{2}$ and $i_{3}$. Then, items $i_{2}$ and $i_{3}$ will be considered as the \yz{adjacent} nodes to the user node $u_{0}$. Afterwards, we use such \zixuan{neighbouring-items} 
as a type of \zx{personalised} graph prompts and \zixuan{call} \zixuan{them} as `hard graph prompts'. Note that we ignore the users' interacted items as \yz{adjacent} nodes \zixuan{when} devising prompts to avoid \io{the possible} over-fitting of the final learned model. 
In addition, we also introduce a different type of \zx{personalised} graph prompts, \io{namely} `soft graph prompts', which consists of a pre-defined number of randomly initialised embedding vectors.
\zx{\io{Overall},  we define both \io{the} hard graph prompts and \io{the} soft graph prompts as the personalised graph prompts of a given user.}
Finally, such personalised graph prompts will be leveraged to assist the graph encoder that is pre-trained from the source domain to improve the recommendation efficiency and effectiveness. In particular, to \io{guide} the pre-trained model in learning \zixuan{richer} 
user/item representations, we embrace the \zixuan{use of the} contrastive learning technique in the pre-training stage. 
\zixuan{To further enhance the model tuning in the target domain, we combine the contrastive learning loss and \zixuan{the} BPR loss} in a multi-task manner. 
\zx{In next section, we introduce the rationale \zx{and motivation for leveraging the} personalised graph prompts in the tuning stage.}



\subsection{Personalised Graph Prompts}\label{cpgp}

In this section, we first justify 
the equivalence between Graph Neural Networks and \zixuan{the} transformer to explain the rationale of introducing our \ziy{personalised} graph \zixuan{prompts}.
Then, we further describe our proposed \zixuan{graph prompts} and \io{illustrate \zixuan{them}} with GNN recommender models. 
\yzx{Furthermore, we explain the rationale behind \yo{the incorporation of} graph prompts in cross-domain recommendation, emphasising the benefits and implications of this approach within the context of recommendation systems.}

\subsubsection{Comparing a Transformer and GNNs}\label{sec:eqv}
\noindent\textbf{Transformer:} The transformer architecture consists of a composition of transformer layers~\cite{vaswani2017attention}. Each transformer layer has two parts: a self-attention module and a position-wise feed-forward
network (FFN). Let $h^{\ell}$ denote the input of the self-attention module on a certain layer. 
\zixuan{The propagation rule of \io{the} transformer}
updates the hidden feature $h$ at position $i$ of a sentence $S$ from layer $\ell$ to layer $\ell+1$ as follows:



\begin{equation}
\label{eqn:trans}
h_{i}^{\ell+1}=\sum_{j \in \mathcal{S}} w_{i j}\left({W}_{V}^{\ell} h_{j}^{\ell}\right)
\end{equation}

\begin{equation}
\label{eqn:att}
w_{i j}=\operatorname{softmax}_{j}\left({W}_{Q}^{\ell} h_{i}^{\ell} \cdot {W}_{K}^{\ell} h_{j}^{\ell}\right)
\end{equation}
where $j \in S$ denotes the set of words in the sentence and \zx{${W}_{Q},{W}_{K},{W}_{V}$} are \yz{learned} linear weights. In natural language processing,
\zixuan{the transformer sums} over all the words in a sentence and outputs the next hidden feature by applying {a} weighted summation on the values.

\noindent \textbf{Graph Neural Network:} Modern GNNs~\cite{kipf2016semi,velivckovic2017graph} follow a learning schema that iteratively updates the representation of a node by aggregating \zixuan{the} representations of its first or higher-order neighbours.
Let $h^{\ell}$ denote the representation of a node $v_{i}$ at \zixuan{the} $l$-th layer.
GNNs update the hidden features $h$ of node $i$ at layer $\ell$ via a non-linear transformation of the node's own features added to the aggregation of features from each neighbouring node $j \in N(i)$ :


\begin{align}
\label{eqn:gnn}
    &h_{i}^{\ell+1}=\sigma\left({W}_{U}^{\ell} h_{i}^{\ell}+\sum_{j \in \mathcal{N}(i)} w_{ij}\left({W}_{V}^{\ell} h_{j}^{\ell}\right)\right)
\end{align}
where \zx{${W}_{U}, {W}_{V}$} 
are \yz{learned} weights of the GNN layer and $\sigma$ is a \zixuan{non-linear transformation}. 
\zixuan{GNNs exploit high-order connectivity by summing} 
over the \final{feature vectors $h_{j}^{\ell}$} from \zixuan{the} \yz{adjacent} nodes with $w_{ij}$  set to 1 and output the next hidden feature. \yz{Given} $w_{ij}$ is a \yz{learned} weight that is dependent on nodes $i$ and $j$, 
\ziy{the} GNN layer 
can be interpreted as a layer in a single head-based GAT~\cite{velivckovic2017graph} and is \zixuan{similar} to the calculation of the hidden features in a transformer layer (see Equation~\eqref{eqn:trans}). 

To conclude, \zixuan{a} transformer sums over all words in a sentence and \yz{a} GNN sums over the local neighbourhood. 
Moreover, \zixuan{a} transformer uses self-attention to have a weighted sum for feature transformation \yz{since} self-attention can be seen as \yz{the inference of} \zixuan{a} soft adjacency \yz{matrix}. 
\zixuan{Indeed},
compared to the transformer architecture, \yz{a} GNN can be considered as a simple transformer that applies a linear-weighted sum on a randomly permuted sentence, since the neighbouring nodes compose a `sentence' and \yz{the} `words' are in random order without the positional embeddings.
Inversely, \zixuan{the} transformers can also be viewed as a special case of GNNs on a fully connected graph of words~\cite{joshi2020transformers}.



\subsubsection{Prompt-tuning in a Transformer and GNNs}\label{sec:ptgnn}
\io{Equation}~(\ref{eqn:trans}) and \io{Equation}~(\ref{eqn:att}) describe the dot-product attention mechanism, which provides a weighted summation over the words in a sentence.
Here, we derive an equivalent form of \io{Equation}~(\ref{eqn:trans}) as follows:

\begin{equation}
\begin{aligned}
h_{i}^{\ell+1} &=\operatorname{Attn}\left(h_{i}^{\ell}{W}_{Q}, h_{j}^{\ell}{W}_{K}, h_{j}^{\ell}{W}_{V}\right) \\
&=\operatorname{Attn}\left({Q}{W}_{Q}, K{W}_{K}, V{W}_{V}\right)
\end{aligned}
\end{equation}
where ${W}_{Q}, {W}_{K}, {W}_{V}$ are \yz{learned} weights on the \zx{layer input} Q 
and \zixuan{the} key-value pairs.

The mechanism of prompt-tuning changes 
the attention module through
prepending the tunable vectors to the original attention keys and values at every layer~\cite{li2021prefix,he2021towards}. Specifically, two sets of prefix vectors ${P}_{K}, {P}_{V} \in \mathbb{R}^{l \times d}$ are concatenated with the original attention key ${K}$ and value ${V}$. Here, we provide an alternative view of prompt-tuning in a transformer:

\begin{equation}
\begin{aligned}
\label{eqn:prefix}
h_{i}^{\ell+1} &=\operatorname{Attn}\left(QW_{Q}, \left({P}_{K}\parallel K{W}_{K}\right), \left({P}_{V} \parallel V{W}_{V}\right)\right)\\
&=\operatorname{softmax}\left(QW_{Q}\left(P_{K} \parallel K{W}_{K}\right)^{\top}\left[\begin{array}{c} P_{V} \\ V{W}_{V} \end{array}\right]\right) \\
&=(1-\lambda) \operatorname{softmax}\left(QW_{Q}W_{K}^{\top}K^{\top}\right)VW_{V} +\lambda \operatorname{softmax}\left(QW_{Q} P_{K}^{\top}\right) P_{V} \\
&=(1-\lambda) \operatorname{Attn}\left(QW_Q, KW_{K}, VW_{V}\right) +\lambda \operatorname{Attn}\left(QW_{Q}, P_{K}, P_{V}\right) 
\end{aligned}
\end{equation}
where $\parallel$ is the concatenation operator, \yz{and} $\lambda$ is a scalar that represents the sum of normalised attention weights on the prefixes:
\begin{equation}
\lambda=\frac{\sum_{i} \exp \left({Q} {W}_{Q} {P}_{K}^{\top}\right)_{i}}{\sum_{i} \exp \left({Q} {W}_{Q} {P}_{K}^{\top}\right)_{i}+\sum_{j} \exp \left({Q} {W}_{Q}{W}_{K}^{\top}K^{\top}\right)_{j}}
\end{equation}
Note that the first term in \io{Equation}~(\ref{eqn:prefix}), $\operatorname{Attn}\left(QW_Q, KW_{K}, VW_{V}\right)$, is the original attention without prefixes, whereas the second term is \zixuan{the} \zixuan{attention modification term}, 
which changes the attention module through linear interpolation on the original output of a transformer:
\begin{equation}
h_{i}^{\ell+1} \leftarrow(1-\lambda) h_{i}^{\ell+1}+\lambda \Delta h_{j}^{\ell+1}, \quad \Delta h_{j}^{\ell+1}=\operatorname{softmax}\left({Q} {W}_{Q} {P}_{K}^{\top}\right) {P}_{V}
\end{equation}

Inspired by the success of prompt-tuning methods, we prepend {the} mixture node embeddings to be the item-wise graph {prompts} in the CDR scenario.
Specifically, we append such node embeddings as prompts to the output of \zixuan{the} pre-trained GNNs to update the learned features of nodes. \yz{These prompts further incorporate additional items or nodes, referred to as neighbouring-items or continuous learned vectors discussed in Section~\ref{sec:intro}, for a target user. As a consequence, each user node obtains a specific number of $m$ prompts. Next, we update the embeddings of the items using their respective adjacent user nodes.}
When we apply the prompt-tuning to GNNs, we can obtain an updated propagation rule on top of \io{Equation}~(\ref{eqn:gnn}):
\begin{align}
    &h_{i}^{\ell+1}=\sigma\left({W}_{U} h_{i}^{\ell}+(1-\lambda) 
\sum_{j \in N_{(i)}} ({W}_{V} h_{j}^{\ell})
+\lambda \sum_{\yz{r} \in N_{(i)}} ({P_{V}}^{\prime} h_{\yz{r}}^{\ell})\right)
\label{eqn:pmt}
\end{align}
where \yz{$r$} is \io{a} neighbouring-items node \io{such as} $\yz{r} \in N(i)$. \io{ \yz{$r$} is used to act as a new adjacency node of the given node, thereby creating a new edge connection in the graph}; 
$\lambda=\frac{\left|N_{\yz{r}}\right|}{\left|N_{j}\right|+\left|N_{\yz{r}}\right|}$ is \io{an} item-related weight factor.
$\left|N_{j}\right|$ and $\left|N_{\yz{r}}\right|$ denote the first-hop neighbours \io{item} $j$ and \io{the} neighbouring-item $r$, respectively. 
\yz{Similar to $W_{V}$, $P_{V}^{'}$ is a learned weight matrix of the GNN layer.
It is worth noting that during the tuning stage, the second term of Equation~(\ref{eqn:pmt}), 
without ($1-\lambda$), is the output from a pre-trained graph encoder with frozen parameters. 
Furthermore, 
the third term of Equation~(\ref{eqn:pmt}), specifically the learned matrix $P_{V}^{'}$, is the only learned matrix within the context of the target domain $D_{\mathrm{T}}$.}
\zixuan{As a result, tuning only on a small set of parameters} is expected to improve \zixuan{efficiency} in \zixuan{the} tuning stage. 
\yzx{The \yy{number of} tuning \yo{parameters} \yy{of the graph encoder}
is determined by considering the \yz{learned} parameters associated with the number of both hard and soft prompt nodes.
\yy{These nodes are aggregated as supplementary nodes within the graph encoder, and their combined parameters \yo{hence} contribute to the total number of \yz{the learned} parameters.}}
As such,
\zixuan{we argue that} the resulting GNN model using the \zx{personalised graph prompts}
is capable of \yy{efficiently} leveraging the useful knowledge from the pre-trained GNN models, \yy{while \yy{also} \yy{reducing} \yo{the} tuning parameters required during \yo{the} tuning stage.} 
\subsubsection{Hard Prompts and Soft Prompts.}
\yy{In this section, we provide a comprehensive overview of both hard and soft prompts in our PGPRec framework, detailing their selection, initialisation, and integration within the framework.}
\\
\yzx{
\noindent\textbf{Hard Prompts.}
Informative personalised prompts necessitate \yo{the inference of} correlations among items.
\yy{Indeed,}
\yo{the implementation of} a straightforward yet efficient 
approach for determining item correlations is essential for generating personalised hard prompts that enable sufficient learning within a target domain.
\yz{Following~\cite{liu2021contrastive}, we introduce a model-based method to infer correlations between the interacted items and their corresponding neighbouring items within the target domain. Liu et al.~\cite{liu2021contrastive} suggested that a model-based correlation measurement is more desirable than alternative methods such as relying solely on the user frequency counts of the given items.} 
\yz{Since the item representations are jointly learned with the graph encoder during the prompt-tuning phase, this correlation score is inherently model-based.}
\yz{In this work, we use the dot-product to measure the correlation between items.}
\yz{Let $\mathbf{e}_\mathbf{j}$ be the representations of all the $\mathbf{j}$ items that a given user has interacted with, and similarly $\mathbf{e}_\mathbf{r}$ denotes
the representations of all of the neighbouring-items $\mathbf{r}$ in the target domain. It follows that the model-based correlation score can be defined as follows:}
\begin{align}
\operatorname{Cor}_{\mathrm{e}}(\mathbf{j}, \mathbf{r})=\mathbf{e}_\mathbf{j} \cdot \mathbf{e}_\mathbf{r}
\end{align}
\yy{In our proposed PGPRec framework}, we \yz{select} the top-$m$ \yz{items from the} neighbouring-items $\mathbf{r}$, \yy{determined by the correlation score}, as hard prompts where $m$ represents the number of hard prompts \yy{integrated within the} framework.
\yz{The hard prompts, which originate from the actual nodes within the graph and which are updated independently, directly affect the total number of parameters. As mentioned in Section~\ref{sec:ptgnn}, the number of these hard prompts is pre-determined based on the top-$m$ correlation score, leading to a consistent parameter scale.}
This \yy{selection} approach ensures that the most relevant and informative neighbouring items are used for generating \yo{the} hard prompts, \yo{thereby} enhancing the overall recommendation performance in the target domain.}
\\
\noindent\textbf{Soft Prompts.}
\yzx{Different from \yo{the} hard prompts, \yo{the} soft prompts adopt a flexible and adaptive approach to capture the underlying item relationships within the target domain. 
\yy{In particular, \yo{the} soft prompts are randomly initialised and subsequently embedded into trainable vectors.} 
\yz{
These soft prompts serve as auxiliary nodes that establish connections with the target users within the target domain. This process facilitates the collection of contextual information from the user-item interactions, thereby enabling our PGPRec framework to generate more effective user/item representations. As a consequence, the number of parameters increases proportionally to an increase in the number of soft prompts.}
Furthermore, since the soft prompts are trainable, they can adapt to the data during the training process, \yy{thereby} aligning the user/item embeddings with the underlying structure of the target domain data. As such, this adaptability results in smoother user/item representations, \yo{since} the soft prompts can better capture more fine-grained user/item embeddings compared to \yo{the} hard prompts.}


\subsubsection{Why Prompt-tuning in GNNs?}
\yzx{\yy{Applying prompt-tuning} in GNNs offers several \yy{potential} advantages for cross-domain recommendation tasks, such as improved tuning efficiency, \yo{the} provision of \yy{additional collaborative signals} for the target domains, and \yo{an} \yy{easier} deployment in large-scale graphs.
\textbf{(1)~Tuning efficiency:} 
Tuning a \yy{smaller set} of parameters within GNNs using both hard and soft prompts is essential, as it allows for the customisation of the model, enabling it to capture the unique characteristics specific to the target domain. 
As illustrated in Equation~(\ref{eqn:pmt}), prompt-tuning in GNNs leverages pre-trained graph encoders and concentrates on a limited set of parameters by tuning solely with the personalised prompts. Subsequently, the obtained embedding is combined with that generated by the pre-trained graph encoder, 
\yy{\yo{therefore} leveraging the previously pre-trained knowledge while maintaining \yo{the} model efficiency.}
\textbf{(2)~Additional collaborative signals:} Prompt-tuning in GNNs offers a flexible solution capable of incorporating supplementary collaborative signals for the target domain~\cite{sun2022gppt,fang2022prompt}. \yy{In our proposed PGPRec framework,} \yo{the} personalised graph prompts, which include both \yo{the} hard and soft prompts, \yy{can indeed} serve as auxiliary side information \yy{for} the target domain. These prompts can be tailored to enhance \yo{the} recommendation performance in specific cases, 
\yo{thereby likely} improving the accuracy and effectiveness of the cross-domain recommender system.
\textbf{(3)~\yy{Easier} deployment:} Applying personalised graph prompts on \yy{adequately pre-trained} GNNs \yy{leads to less} parameters to store~\cite{sun2022gppt}, rendering it a \yy{more} convincing option for large-scale graph applications where \yo{the} training \yo{of}  GNNs necessitates significant computational resources and memory. \yy{In our proposed PGPRec framework,} by freezing \yo{the} GNNs once they are pre-trained and tuning \yo{the} graph prompts in an efficient and scalable manner, \yy{as described in Section~\ref{sec:ptgnn}}, 
PGPRec can \yy{potentially} be deployed for various applications without the need for extensive training.}
\subsection{Contrastive Learning in PGPRec}\label{sec:cl4rec}

   
Unlike the existing approaches, we adopt the Contrastive Learning (CL) scheme~\cite{oord2018representation} \zixuan{to ensure \zx{an effective} training} in both the pre-training and fine-tuning phases.
In the pre-training stage, we employ contrastive learning to optimise the graph encoder with the data from the source domain $D_{\mathrm{S}}$. 
Specifically, the \zixuan{contrastive} pre-training \zixuan{stage} comprises three main components: 1) a graph augmentation tool, which builds distinct sub-graphs for generating the representation variants of identical nodes, 2) a graph encoder to model the developed sub-graphs, and 3) a corresponding contrastive loss function for \zixuan{the} graph encoder optimisation. 
On the other hand, to further enhance the recommendation \io{performance} in \io{the} tuning stage, we \zx{also leverage} 
the contrastive learning technique and introduce a joint-learning loss that \zixuan{comprises} both \zixuan{the} contrastive loss and the commonly used BPR~\cite{rendle2009bpr} loss. \ziy{To better illustrate the training process of PGPRec, Algorithm 1 \io{provides the training pseudo-code}.}

\subsubsection{Contrastive Pre-training of PGPRec}\label{sec:cl}
We \final{now} describe the contrastive learning component in our PGPRec framework.
Following~\cite{wu2021self}, we generate two augmented sub-graphs via \io{the} edge dropout \ziy{technique} and feed them into the graph encoder $f$. 
\io{As \io{mentioned} in Section~\ref{sec:frame}, PGPRec is a model-agnostic framework \io{that} is flexible \io{to \yy{accommodate}} the commonly used GNN models.} In this paper, we adopt a special case of LightGCN~\footnote{\io{Since} the only trainable model parameters in LightGCN are the user/item embeddings at the 0-th layer, LightGCN cannot transfer the structure knowledge with the feature transformation matrices from the source domain. Hence, following~\cite{yang2022stam}, we adopt LightGCN with the attention aggregator to explicitly learn transferable knowledge. }~\cite{he2020lightgcn} as the graph encoder, which is a simple yet efficient GNN recommender~\cite{mao2021ultragcn}. 
\ziy{Specifically, we use the edge dropout augmentation with the best \io{reported performance} in SGL~\cite{wu2021self} \zx{which is an effective graph contrastive learning method. Then, we can obtain two augmented sub-graphs as follows:}}
\begin{equation}
    \label{eqn:aug}
    \mathcal{G}_{S'} = (\mathcal{V}, M \odot \mathcal{E}), \  \ \mathcal{G}_{S''} = (\mathcal{V}, {M'} \odot \mathcal{E})
\end{equation}
\ziy{where $\mathcal{V}$ is the node set, \io{while} $M$ and $M'$ are \zx{two different} masking vectors 
on \io{the} edge set $\mathcal{E}$.}

By using the augmented sub-graphs and a graph encoder, we can generate different embedding representations of an identical node (i.e.\ positive pair). For example, consider a given node $q$. \zixuan{We} can obtain its embeddings $e_{q}$ and $e_{q'}$ after encoding two sub-graphs with the graph encoder $f$. Then, to obtain a negative pair \zixuan{of sub-graphs}, we apply an in-batch \zixuan{random} sampling strategy (e.g.\ a pair of node representations $e_{q}$ and $e_{k}$ of two different nodes $q$ and $k$). 
\zixuan{By contrasting the positive and negative pairs}, we expect \zx{that} the resulting user/item representations of users/items can effectively improve the recommendation performance when applied to the target domain. 
\zx{As such,} the learned feature embeddings from the source domain \zxx{allow} the model to start from a better initialisation point 
and provide promising results after a light fine-tuning~\cite{gouk2020distance} in various target domains.

Next, after obtaining the positive and negative pairs, we follow SimCLR~\cite{chen2020simple} to \zixuan{generate a} better representation via data augmentations
and adopt the contrastive loss, InfoNCE~\cite{oord2018representation}, to maximise the agreement of \zixuan{the} positive pairs and minimise that of the negative pairs:
\begin{equation}
\begin{aligned}
    \mathcal{L}^{user}_{cl} &= -\log \frac{\exp \left(\mathbf{e}_{q}^{\top} \mathbf{e}_{q'} / \boldsymbol{\tau}\right)}{\sum_{i=1}^{n} \exp \left(\mathbf{e}_{q}^{\top} \mathbf{e}_{k} / \boldsymbol{\tau}\right)} 
\label{eqn:cl}
\end{aligned}
\end{equation}
where $\tau$ is a hyper-parameter that adjusts the dynamic range of the resulting loss value.
Analogously, we obtain the contrastive loss of the item side $\mathcal{L}^{item}_{cl}$. 
Combining these two losses, we obtain an objective function for the contrastive task as \io{follows}: 
\begin{equation}
    \mathcal{L}_{cl} = \mathcal{L}^{user}_{cl} + \mathcal{L}^{item}_{cl}
    \label{eqn:uicl}
\end{equation}

\subsubsection{Contrastive Tuning of PGPRec}\label{sec:tunpgprec}
Once \zx{we obtain} the pre-trained GNN model from the source domain, we transfer the learned model to the target domain. Recall the discussion in Section~\ref{sec:frame} \zixuan{where} we introduced two variants of \zx{the personalised} graph prompt, \zixuan{namely} \io{the} soft and hard graph prompts \zixuan{in order} to improve the recommendation efficiency and effectiveness in a CDR setting.
\zixuan{Differently} from the pre-training phase, we optimise both a pairwise ranking task objective and a contrastive learning objective $\mathcal{L}_{cl}$ \zixuan{to further \zixuan{integrate the} prediction signals during \io{the} tuning phase:} 
\begin{equation}
\begin{aligned}
    \mathcal{L} = &\mathcal{L}_{rec} + \lambda_{1}\mathcal{L}_{cl} + \lambda_{2} \left \|\Theta  \right \|^{2}_{2}
    \\&\operatorname{where} \mathcal{L}_{rec} = \sum_{(u,i,j)\in D_{s}} -\text{\final{log}}\sigma (\boldsymbol{e_{u}}^{\top}(\boldsymbol{e_{i}} - \boldsymbol{e_{j}}))
    \label{eqn:mmloss}
\end{aligned}
\end{equation}
where \yzx{$\mathcal{L}_{rec}$} is \yzx{the Bayesian Personalised Ranking (BPR) loss \cite{rendle2009bpr}}, $\boldsymbol{e_{u}}$ is the user embedding, $\boldsymbol{e_{i}}$ denote\xiw{s} the positive item embedding and ${y}_{ui}$ is \xiw{the} ground truth value, $D_{s}\ =\{(u,i,j)|(u,i)\in R^{+},(u,j)\in R^{-}\}$ is the set of the training data, $R^{+}$ indicates the interacted user-item pairs and $R^{-}$ indicates user-item pairs from \io{the} users' unseen items. Moreover, $\sigma(\cdot)$ is the sigmoid function, $\Theta$ is the set of model parameters in \yzx{$\mathcal{L}_{rec}$} {while} $\lambda_{1}$ and $\lambda_{2}$ are hyper-parameters to control the strengths of \io{the} CL and $L_{2}$ regularisation, respectively. 

\IncMargin{1em} 
\begin{algorithm}
\caption{\ziy{The overall training process of \io{the} PGPRec framework}}\label{alg:train}
    \SetAlgoNoLine 
    \SetKwInOut{Output}{\textbf{Output}}
    \SetKwInOut{Return}{\textbf{return}}
    \KwIn{
    \\ 
    \   \ \   \  \   \ \   \ \   \ \   \ The user-item interaction graphs $\mathcal{G}_{S}$ and $\mathcal{G}_{T}$\;\\
    \   \ \   \  \   \ \   \ \   \ \   \ The batch size $B$\;\\
    \   \ \   \  \   \ \   \ \   \ \   \ The hyper-parameters $\lambda_{1},\lambda_{2}$\;\\
    \   \ \   \  \   \ \   \ \   \ \   \ The ID representations of neighbouring-item $\boldsymbol{e_{p_h}}$ as hard \io{prompts}\;\\}
    \KwOut{\\
    \   \ \   \  \   \ \   \ \   \ \   \ The final user/item embeddings $e_{u}$, $e_{i}$\;\\}
    \BlankLine
    Initialise epoch $t = 0$\; 
    Initialise the model parameters $\Theta$ with the default Xavier distribution\; 
\tcp{Pre-training parameters in the source domain $\mathcal{D}_{S}$ }
    \Repeat
        {\text{convergence}}
        {$t = t + 1$\;
        Obtain \io{the} original ID representations $\boldsymbol{e_{u}}$, $\boldsymbol{e_{i}}$ for each user $u$ and item $i$ based on $\mathcal{G}_{S}$ \;
        Obtain \io{the} sub-graphs $\mathcal{G}_{S'}$ and $\mathcal{G}_{S''}$ with \io{a} graph perturbation on $\mathcal{G}_{S}$ with Eq.(\ref{eqn:aug})\;
        Obtain \io{the} perturbed user/item representations $e_{q}$, $e_{q'}$ based 
        on sub-graphs $\mathcal{G}_{S'}$ and $\mathcal{G}_{S''}$ with Eq.(\ref{eqn:gnn}) \;
        Calculate \io{the} InfoNCE \yzx{losses $\mathcal{L}^{user}_{cl}$ and $\mathcal{L}^{item}_{cl}$} with $e_{q}$ and $e_{q'}$, according to Eq.(\ref{eqn:cl})\;
        Calculate \io{the} \yzx{combined} InfoNCE loss $\mathcal{L}_{cl}$ with \yzx{Eq.(\ref{eqn:uicl})} \;
        Backpropagation and update model parameters $\Theta$ with $\mathcal{L}$\;
        }
\tcp{Tuning parameters in the target domain $\mathcal{D}_{T}$}
    Initialise epoch $t = 0$\;
    \Repeat
        {\text{convergence}}
        {$t = t + 1$\;
        Obtain \io{the} original ID representations $\boldsymbol{e_{u}}$, $\boldsymbol{e_{i}}$ for each user $u$ and item $i$ based on $\mathcal{G}_{T}$ \;
        Acquire the hard prompt representations $\boldsymbol{e_{p_h}}$ through neighbouring-items\;
        Acquire the soft prompt representations $\boldsymbol{e_{p_s}}$ with random initialised embedding vectors\;
        Encode the user/item representations $\boldsymbol{e_{u}}$, $\boldsymbol{e_{i}}$ through \io{the} pre-trained model parameters $\Theta$\ with Eq. (\ref{eqn:gnn})\;
        Combine \io{the} embeddings of $\boldsymbol{e_{p_h}}$ and $\boldsymbol{e_{p_s}}$ into the user/item representations $\boldsymbol{e_{u}}$, $\boldsymbol{e_{i}}$ with Eq. (\ref{eqn:pmt})\;
        Calculate \io{the} BPR loss $\mathcal{L}_{BPR}$ with Eq. (\ref{eqn:mmloss})\;
        Obtain \io{the} perturbed user/item representations $e_{q}$, $e_{q'}$ based 
        on sub-graphs $\mathcal{G}_{T'}$ and $\mathcal{G}_{T''}$ with Eq. (\ref{eqn:aug}) \;
        Calculate \io{the} InfoNCE loss $\mathcal{L}_{cl}$ for contrastive learning with $e_{q}$ and $e_{q'}$, according to Eq. (\ref{eqn:cl})\;
        Calculate \io{the} joint learning loss $\mathcal{L}$ with Eq. (\ref{eqn:mmloss}) \;
        Backpropagation and update \io{the} model parameters $\Theta$ with $\mathcal{L}$\;
        }
    \KwRet {the final user/item embeddings $e_{u}$, $e_{i}$ in the target domain $\mathcal{D}_{T}$} \:
\end{algorithm}
\DecMargin{1em}


\section{Experiments}\label{sec:exp}
To demonstrate the \zixuan{efficiency} of PGPRec and \io{provide evidence} 
for its effectiveness 
, we conduct experiments to answer the following \ziy{four} research questions:


\noindent \textbf{RQ1}: How \zixuan{do} the PGPRec framework perform in \zixuan{cross-domain recommendation} compared with \zixuan{existing} baselines?
    
\noindent \textbf{RQ2}:
\zixuan{How do different hard graph prompts and soft graph prompts and their combination impact the recommendation performance?}

\noindent \textbf{RQ3}: \zx{Are the personalised} graph prompts more efficient \zxy{than the fine-tuning} in cross-domain recommendation? 

\noindent \textbf{RQ4}: \ziy{\io{Are} the personalised graph prompts \io{effective} in a cold-start scenario?}

\subsection{Experimental Setup}
\subsubsection{Datasets}\label{sec:dataset}

To evaluate the effectiveness of our PGPRec framework, we conduct experiments on four \textit{Amazon Review datasets}~\footnote{\url{https://jmcauley.ucsd.edu/data/amazon/}}, \zixuan{namely} Electronics (Elec for short), \zx{Cell} Phones (Phone for short), Accessories, Sports and Outdoors (Sport for short) \& Clothing Shoes and Jewelry (Cloth for short). \zixuan{Since \zixuan{the} Amazon Review datasets include information from items in different domains, \zixuan{they are} suitable for evaluating cross-domain recommendation~\cite{wang2020cross}}. In particular, \zixuan{these datasets are considered} into 4 pairs of source-target datasets as shown in Table~\ref{tab:persp_cdr}. Each pair of datasets \zixuan{shares} the common \zixuan{users} between the source and target \zixuan{datasets}. 
The \zixuan{exact} statistics of \zixuan{the used pairs of} datasets are listed in Table~\ref{tab:persp_cdr}. 
By comparing the statistics across all datasets in Table~\ref{tab:persp_cdr}, it is clear that \zixuan{each source-target \zixuan{dataset} pair exhibits} a similar density level.
\zixuan{When processing the datasets, we transform the ratings into \final{1 or 0}, indicating whether the user has rated the item \final{or not}. 
\io{Following} ~\cite{liu2020cross}, \io{we denote those users that have more than 5 interactions in a given dataset as the regular users while the cold-start user are those users with less than five ratings.}}


\begin{table}[!h] 
\begin{center}
\small
\caption{Statistics of \zixuan{the used} Amazon Review datasets. \io{The column 'Users'} donates the number of overlapping users and \io{the column 'Cold-start users'} donates the number of cold-start \io{users} in \io{the} test set.}
\begin{adjustbox}{width=0.6\linewidth}
\begin{tabular}{c|c c c c c c}
\cline{1-5}
\hline
Dataset \ & Users & Cold-start users & Items & Interactions & Density\\
\hline
Elec & 12,739  &  2,014 & 36,183 & 465,545 & 0.101\%  \\ 
Phone & 12,739  & 2,014 & 12,772  &  178,973 & 0.110\%  \\ 
\hline
Sports & 6,755  & 893 & 31,514 & 93,666 & 0.044\% \\ 
Cloth & 6,755  & 893 & 35,131  &  87,805 & 0.037\% \\
\hline
\hline
Sports & 4,017  & 617 & 19,396 & 52,202 & 0.067\% \\ 
Phone & 4,017  & 617 & 12,517  &  40,225 & 0.080\% \\ 
\hline
Elec & 11,611  & 1,803 & 49,730 & 202,095 & 0.035\% \\ 
Cloth & 11,611  & 1,803 & 47,265  &  137,198 & 0.025\% \\
\hline
\hline
\end{tabular}
\end{adjustbox}
\label{tab:persp_cdr}
\end{center}
\end{table}

\subsubsection{Evaluation Metrics}\label{sec:eva}
In this work, we adopt two widely used evaluation metrics to evaluate the performance of our PGPRec \io{framework as well as the used} baselines.
Specifically, suppose there is a set of items to be ranked. Given a user $u$, let $\hat{R}(u)$ represent a ranked list of items that an algorithm produces, and \io{let} $R(u)$ represent a ground-truth set of items that user $u$ has interacted with. For \io{the} top-$k$ recommendation \io{task}, only \io{the} top-ranked items are important to consider. \zx{Therefore, we list the used metrics below:}
\begin{itemize}
    \item Recall at top-$k$ positions: Recall is a metric for computing the fraction of relevant items out of all relevant items and is defined as \io{follows}:
    \begin{equation}
    \text { Recall@k }=\frac{1}{|{U}|} \sum_{u \in {U}} \frac{|\hat{R}(u) \cap R(u)|}{|R(u)|} \text {, }
    \end{equation}
    where $\hat{R}(u)$ is \io{the} ranked list of items generated by the recommenders and $|\hat{R}(u)|$ is the size of the ranked list $\hat{R}(u)$, \io{$R(u)$} is the ground-truth set of items that user $u$ has interacted with and $|R(u)|$ represents the size of \io{the} item set $R(u)$, ${U}$ denotes the user set with size $|{U}|$. Here, $|\hat{R}(u)|=k$.
    
    \item \yo{Normalised} Discounted Cumulative Gain at top-$k$ positions:
    \yo{Normalised} Discounted Cumulative Gain (NDCG) is a metric \zx{devried from Discounted Cumulative Gain (DCG)~\cite{jarvelin2002cumulated}}, which takes the positions of \io{the} correct recommended items into consideration, where \io{the} positions are discounted logarithmically~\cite{ricci2011introduction}. \io{NDCG} accounts for the position of the hit by assigning higher scores to hits at \io{the} top ranks~\cite{he2017neural} and is defined as \io{follows}:
    \begin{equation}
   \text { NDCG@k }=\frac{1}{|\mathcal{U}|} \sum_{u \in \mathcal{U}} \frac{1}{Z} \sum_{x=1}^{|\hat{R}(u)|} \frac{2^{I\left(\hat{R}_x(u) \in R(u)\right)}-1}{\log _2(x+1)}
    \end{equation}
    where $\hat{R}_x(u)$ \final{denotes} for the item recommended \io{at the} $x$-th position, and $Z$ is a \io{normalised} factor, which denotes the ideal value of $DCG$ 
    given $R(u)$. 
\end{itemize}
We measure \io{the recommendation} effectiveness on four Amazon Review datasets described in Section~\ref{sec:dataset} in terms of Recall and NDCG calculated to rank depth 10.
\final{To evaluate parameter-efficiency, we count the parameters associated with $m$ graph prompts $h_{r}$ and the weighted matrix $P_{V}$, as defined in Equation~\eqref{eqn:pmt}, and compare these to the parameter \final{counts} of the pre-trained GNN model.}
\final{Additionally, we evaluate the time-efficiency by measuring the total tuning time, the time per epoch, and the number of epochs needed for convergence as per early-stopping criteria.}
Moreover, we follow the experimental setup in ~\cite{cui2020herograph} and randomly split a given dataset into training, validation, and testing sets with \zixuan{an} 8:1:1 ratio.
\io{For statistical significance comparisons} with the baselines, we use the two one-sided \io{equivalence} test (p < 0.05) and apply the Holm–Bonferroni multiple testing correction, as per best practices in information retrieval~\cite{sakai2021fuhr}.
Moreover, we calculate the used metrics \io{based} on the generated ranking list of items and report the average score over all test users.

\subsubsection{Baselines}\label{sec:baseline}
To evaluate the effectiveness of our \zixuan{proposed approach}, we compare PGPRec with the following \zixuan{existing} state-of-the-art baselines. 
\ziy{The baselines can be roughly \io{categorised} into four groups: (1) NGCF, LightGCN and SGL are single domain collaborative filtering methods, (2) CMF, CoNet and PPGN are joint-learning methods for cross-domain recommendation, (3) NGCF, LightGCN, and SGL are graph pre-training and fine-tuning methods, (4) our PGPRec is a graph pre-training and prompt-tuning method. Table~\ref{tab:persp} \yo{summarises} the baselines \io{as well as} PGPRec across different aspects, \yy{\yo{namely} single-domain, joint-learning, and pre-training followed by fine-tuning.}}
\zx{\io{Below}, we provide a brief description of all \io{the} used baselines.}
\\
\textbf{CMF~\cite{tang2016cross}:} \zixuan{This} is a joint-learning approach, \io{which} factorises matrices of \zixuan{multiple domains} simultaneously by sharing the user latent factors. \zixuan{CMF} first jointly learns on two domains and then optimises the target domain.
\\
\textbf{CDMF~\cite{loni2014cross}:}
\yzx{This is also a joint-learning approach. Different from CMF, CDMF leverages factorisation machines to model interactions between users and items across different domains.
CDMF learns shared latent factors across these domains by representing user-item interactions as feature vectors and applying factorisation machines to model their relationships.
\yo{\yy{S}uch an approach} enables CDMF to capture more complex interaction patterns between users and items, \yy{as well as to} share latent factors in cross-domain recommendations.}
\\
\textbf{CoNet~\cite{hu2018conet}:} \zixuan{This is also} a joint-learning method, which transfers knowledge across domains by \zixuan{leveraging} cross-connections between the feed-forward neural networks.
\zx{Different from CMF, \zixuan{CoNet} optimises both source and target domains with a joint-learning objective.}
\\
\textbf{PPGN~\cite{zhao2019cross}:} \zixuan{PPGN} is a joint-learning graph method, which fuses the interaction information of the two domains into a graph, and shares the features of users learned from the joint interaction graph by stacking multiple graph convolution layers. Finally, it inputs the learned embeddings to the domain-specific MLP structure to learn the matching function.
\\
\textbf{NGCF~\cite{wang2019neural}:} NGCF is a single-domain GCN-based model. It first captures the high-order connectivity information in the embedding function by stacking multiple embedding propagation layers. \zixuan{Next, it concatenates} the obtained embeddings and uses the inner product to make predictions. \zxx{In order to examine the effectiveness of single domain GNNs as well as the fine-tuning of GNNs in Cross-Domain Recommendation (CDR),} \zxy{we perform two different training strategies with NGCF: \final{(1)} single domain training where we directly train in the target domain (denoted by NGCF$_{SD}$) \io{-- this is the typical training strategy in single-domain recommenders}, and \final{(2)} pre-training in a source-domain then fine-tuning in the target domain (denoted by NGCF$_{PT}$).}  
\\
\textbf{LightGCN~\cite{he2020lightgcn}:} This is also a single-domain GCN-based model \zixuan{that} \zxy{\yo{omits} \yz{learned} weight matrices} from NGCF. It simplifies the design in the feature propagation component by removing the non-linear activation and the transformation matrices. 
\zx{\io{Similar to} NGCF, we \io{also} perform both \io{the} single domain training (denoted by LightGCN$_{SD}$) \io{and} the "pre-training then fine-tuning'' process (denoted by LightGCN$_{PT}$).}
\\
\textbf{SGL~\cite{wu2021self}:} This is a single-domain CL-based method. With LightGCN as the encoder of \zixuan{the} users/items, it adopts different augmentation operators \zixuan{such as} edge dropout and node dropout, on the pre-existing features of \zixuan{the} user/items. \io{Similar to} \zx{NGCF \io{and} LightGCN}, 
\zx{\io{we} also perform both the single domain training (denoted by SGL$_{SD}$) and \io{the} "pre-training then fine-tuning'' \io{process} (denoted by SGL$_{PT}$) but using an auxiliary contrastive loss \io{as described} in Section~\ref{sec:cl}.}
\\
\textbf{BiTGCF~\cite{liu2020cross}:} \yzx{This joint-learning graph approach \yo{extends} LightGCN for cross-domain recommendation tasks by employing dual linear graph encoders to generate user and item representations in each domain. Subsequently, a feature transfer layer is used to fuse user representations across domains, effectively capturing the underlying relationships and facilitating enhanced recommendations in the target domain.}
\\
\textbf{PGPRec-BPR:} This is a \zixuan{variant} of PGPRec. \zixuan{\zixuan{Differently} from PGPRec, which pre-trains with \zixuan{a} contrastive loss, PGPRec-BPR \zixuan{uses a} BPR loss to optimise the GNN model.}
Moreover, it tunes with \zixuan{the} same joint loss (Equation~\eqref{eqn:mmloss}) \zx{as} PGPRec.
It is a pre-training and prompt-tuning method. Hence, PGPRec-BPR allows us to gauge the added-value of leveraging contrastive learning.

\begin{table}[tb] 
\begin{center}
    \caption{Summary of \io{compared} approaches across different aspects.}
\begin{adjustbox}{width=1\linewidth}
\begin{tabular}{c|c c c c c c c c c}
\cline{1-10}
Method \ & CMF & CDMF & CoNet & PPGN & BiTGCF & NGCF & LightGCN & SGL & PGPRec\\
\hline
Single-domain & $\times$ & $\times$ & $\times$ & $\times$ & $\times$ & \checkmark  & \checkmark & \checkmark & $\times$\\ 
Joint-learning & \checkmark  & \checkmark & \checkmark & \checkmark & \checkmark & $\times$  & $\times$ & $\times$ & $\times$\\ 
Pre-training $\And$ Fine-tuning & $\times$   & $\times$ & $\times$  &  $\times$  & $\times$ & \checkmark  & \checkmark & \checkmark & $\times$\\ 
Pre-training $\And$ Prompt-tuning  & $\times$  & $\times$ &  $\times$  & $\times$ & $\times$ & $\times$  & $\times$ & $\times$ & \checkmark\\ 
\hline
\end{tabular}
\end{adjustbox}
\label{tab:persp}
\end{center}
\end{table}
\subsubsection{Implementation Details and Hyper-parameter Settings}\label{sec:implement}

\zxy{For a fair comparison \final{between our PGPRec framework and the used baselines}, we conduct all experiments on the same machine with a GeForce RTX 2080Ti GPU.}
\zx{Moreover,} we \io{use} the learned parameters at the pre-training stage to initialise the model parameters at the tuning
stage. In the pre-training stage, the dropout rate $\rho$ of nodes and edges are set \io{to} 0.1 and the softmax temperature $\tau$ in \io{the} contrastive learning loss is set to 0.2, which are reported with the best performance \final{of the original SGL paper}~\cite{wu2021self}.
Furthermore, in \io{the} tuning stage, we tune the hyper-parameters \final{of our PGPRec framework} on the validation set. The learning rate is selected from $\left \{10^{-2},10^{-3},10^{-4}  \right \}$. 
For those hyper-parameters unique to PGPRec,  we tune $\lambda_{1}$, $\lambda_{2}$, $\tau$ and $\rho$ within the ranges of ~$\left \{ 0, 0.1,0.2,...,1.0 \right \}$, $\left \{ 0, 0.1,0.2,...,1.0 \right \}$, $\left \{ 0, 0.1,0.2,...,1.0 \right \}$ and $\left \{ 0, 0.1,0.2,...,0.9 \right \}$, respectively.
We adopt two widely used evaluation metrics, \zixuan{namely} Recall@K and NDCG@K to evaluate the performance of top-K recommendations, \zx{which \io{were described} in Section~\ref{sec:eva}}.
We follow~\cite{simmons2017top} \io{and} set K = 10 and report the average performance achieved for all users in the test set. 
The negative items of each user are defined as those having no interactions with the user. 
Following~\cite{he2020lightgcn}, the dimensions of user and item embedding are set \io{to} 64 to achieve \io{a} trade-off between performance and time cost, which is determined by \io{a} grid search in the range of $\left \{ 16, 32, 64, 128, 256 \right \}$. 
\zxy{We tune NGCF$_{PT}$, LightGCN$_{PT}$, SGL$_{PT}$ \io{and PGPRec-BPR} as described in Section~\ref{sec:baseline} on the validation set. \yzx{For the other cross-domain recommendation baselines \yy{corresponding to} joint-learning approaches \yy{(CMF, CDMF, CoNet, PPGN, BiTGCF),} 
}}
we follow the reported optimal parameter settings \zx{by the authors of \io{these} baselines}.
\zxy{For our PGPRec \io{framework}, we tune the hyper-parameters as described above on the validation set.}
\zxy{Moreover,} we adopt the Xavier initialisation to initialise \zixuan{all} the \zixuan{models'} parameters and use \zixuan{the} Adam optimiser for \yo{the} model optimisation with a batch size of 1024. We apply early-stopping during training, terminating the training when the validation loss does not decrease for 50 epochs.

\begin{table*}[tb]
\centering
\caption{Experimental results \zixuan{for} PGPRec and \zixuan{the} \io{used} baselines. The best \zixuan{performance} is highlighted in bold and the second best result is highlighted with \zixuan{an} underline. 
$^{*}$ and $^{\triangle}$ mean p <0.05 in the t-test and TOST test (AP=0.05) compared to the result of PGPRec
with \io{the} \zx{Holm–Bonferroni correction}.
SD/PT are the abbreviations \zixuan{for} Single-Domain and Pre-Training, \zixuan{respectively.}}
\label{tab:comp_cdrs}
\begin{adjustbox}{width=\linewidth}
\begin{tabular}{ccccccccc}
\toprule
\multirow{1}{*}{\textbf{Dataset}} & \multicolumn{2}{c}{Elec-Phone} & \multicolumn{2}{c}{Sport-Cloth} & \multicolumn{2}{c}{Sport-Phone} & \multicolumn{2}{c}{Elec-Cloth}\\ 
\cmidrule(lr){1-1} \cmidrule(lr){2-3} \cmidrule(lr){4-5} \cmidrule(lr){6-7} \cmidrule(lr){8-9}
Methods & Recall@10 & NDCG@10  & Recall@10 & NDCG@10  & Recall@10 & NDCG@10 & Recall@10 & NDCG@10\\
\midrule
CMF & ${0.4015}^{*}$  & ${0.2543}^{*}$  &  ${0.4309}^{*}$ & ${0.2685}^{*}$  &  $0.3348^{*}$ & ${0.1927}^{*}$  &  ${0.3054}^{*}$ & ${0.1836}^{*}$\\
\yzx{CDMF} & ${0.4457}^{*}$  & ${0.2612}^{*}$ & ${0.4561}^{*}$ & ${0.2874}^{*}$ &  ${0.3808}^{*}$ & ${0.2242}^{*}$ & ${0.3524}^{*}$  & ${0.2356}^{*}$\\
CoNet & ${0.4514}^{*}$  & ${0.2696}^{*}$  &  ${0.4912}^{*}$ & ${0.3128}^{*}$ & ${0.3854}^{*}$ &  ${0.2331}^{*}$ & ${0.3546}^{*}$ & ${0.2353}^{*}$\\
PPGN & ${0.4464}^{*}$  & ${0.2629}^{*}$ & ${0.4678}^{*}$ & 0.2930$^{*}$ & ${0.3677}^{*}$  &  ${0.2240}^{*}$ & ${0.3404}^{*}$ & ${0.2184}$\\
\yzx{BiTGCF} & ${0.5064}^{*}$  & ${0.2999}^{*}$ & ${0.5360}^{*}$ & $\underline{0.3315}^{\triangle}$ &  $\underline{0.4688}^{*}$ & ${0.3072}^{*}$ & ${0.4209}$  & ${0.2783}^{*}$\\
\midrule
NGCF$_{SD}$ & ${0.4501}^{*}$  & ${0.2643}^{*}$ & ${0.4889}^{*}$ & ${0.3027}^{*}$ &  ${0.3955}^{*}$ & ${0.2431}^{*}$ & ${0.3878}^{*}$  & ${0.2483}^{*}$\\
LightGCN$_{SD}$ & ${0.4776}^{*}$  & ${0.2993}^{*}$ & ${0.5327}^{*}$ & \textbf{0.3358}$^{\triangle}$ &  ${0.4557}^{*}$ & ${0.3006}^{*}$ & $\underline{0.4261}$  & ${0.2699 }^{*}$\\
SGL$_{SD}$ & ${0.4722}^{*}$  & ${0.2941}^{*}$ & ${0.5296}^{*}$ & ${0.3219}^{*}$ &  ${0.4486}^{*}$ & ${0.2749}^{*}$ & ${0.4012}^{*}$  & ${0.2569}^{*}$\\
NGCF$_{PT}$ & ${0.4665}^{*}$  & ${0.2969}^{*}$ & ${0.4983}^{*}$ & ${0.3253}^{\triangle}$ & ${0.4251}^{*}$  &  ${0.2681}^{*}$ & ${0.3723}^{*}$ & ${0.2359}^{*}$\\
LightGCN$_{PT}$ & ${0.4821}^{*}$  & ${0.3044}^{*}$ & ${0.5196}^{*}$ & ${0.3268}^{\triangle}$ &  ${0.4543}^{*}$ & ${0.2986}^{*}$ & ${0.4137}$  & ${0.2727}^{*}$\\
SGL$_{PT}$ & \underline{0.5071}  & \underline{0.3282}$^{\triangle}$ & \underline{0.5398} & ${0.3119}^{*}$ & \textbf{0.4721} & \textbf{0.3114}$^{\triangle}$ & \textbf{0.4301}$^{\triangle}$ & \textbf{0.2967}$^{*}$\\
\midrule
PGPRec & \textbf{0.5213}  & \textbf{0.3386} & \textbf{0.5678} & {0.3281} & {0.4607}  & \underline{0.3086} & {0.4233} & \underline{0.2854}\\
\midrule
{\%Diff.} & {2.80}\% & {3.17}\%  & {3.27}\% & {- 2.29}\%  & {- 2.41}\% & {- 0.10}\%  & {- 1.58}\% & {- 3.80}\%\\
\bottomrule
\end{tabular}
\end{adjustbox}
\end{table*}


\subsection{PGPRec Effectiveness Evaluation (RQ1)}
\zixuan{Table \ref{tab:comp_cdrs} reports the empirical results of our PGPRec method in comparison to all \yo{of} the baselines,} \zx{which \io{were} described in Section~\ref{sec:baseline}}. 
\yzx{We evaluate our PGPRec framework in comparison \yo{to} three distinct \yo{recommendation approaches}: GNN-based recommenders, single-domain recommenders and joint-learning recommenders.}
\zx{Specifically, we compare our PGPRec framework to the GNN-based recommenders (NGCF, LightGCN, SGL), which are trained in both \zxy{the} single domain (SD for short) setting \io{as well as in} \zxy{the} pre-training (PT for short) setting.}
\yzx{\yo{In addition}, we compare our PGPRec framework to \yo{the} joint-learning methods (CMF, CDMF, CoNet, PPGN, BiTGCF).}
\yzx{Among the evaluated baselines, LightGCN$_{SD}$ generally outperforms \yo{the} joint-learning CDR methods, with the exception of BiTGCF. This observation highlights the \yy{importance} of investigating the nonlinear interaction relationship between users and items through graph neural networks, \yy{as mentioned in Section~\ref{sec:baseline}}. 
Notably, BiTGCF is a \yo{tailored} approach that builds upon LightGCN. 
\yy{The comparison between LightGCN$_{SD}$ and BiTGCF} further supports the effectiveness of \yo{the} GNN-based methods in capturing complex user-item relationships within the cross-domain recommendation context.}
\yy{Furthermore, when comparing SGL$_{PT}$ with NGCF and LightGCN, as well as their single-domain (NGCF$_{SD}$, LightGCN$_{SD}$) and pre-trained variants (NGCF$_{PT}$, LightGCN$_{PT}$), SGL$_{PT}$
exhibits \yo{a} superior performance in 88\% of the cases  \yo{(28 out of 32 instances)}, with a significant difference.}
This result emphasises the advantages of employing contrastive learning during both \yo{the} pre-training and tuning stages, \yo{since} it facilitates the generation of richer user representations, ultimately enhancing the recommendation performance.
\zixuan{Comparing PGPRec} and SGL$_{PT}$, \zixuan{we observe that} PGPRec is competitive \zixuan{and comparable} with SGL$_{PT}$ \zixuan{on} all datasets. \zixuan{In fact, PGPRec \zixuan{has} even a} better performance in \yzx{75\% of the} cases \yo{(3 out of 4 instances)}
\zixuan{\yy{on} the} 
Elec-Phone and Sport-Cloth datasets, \zixuan{\zixuan{thereby} demonstrating the general} effectiveness of personalised graph prompts. However, PGPRec performs worse than SGL$_{PT}$ \zixuan{on the} Sport-Phone \& Elec-Cloth datasets. 
\yzx{This may be due to PGPRec being more sensitive to \yy{differences between domains, such as variations in user-item interaction patterns or preferences,}
compared to SGL-PT. As a result, such sensitivity may affect the PGPRec's 
performance when transferring knowledge between distant domains. This observation warrants further investigation to better \yy{identify} 
the factors influencing PGPRec’s performance in such scenarios. We leave such an investigation to future work.}


Hence, \zixuan{in} answer \zixuan{to} RQ1, \zixuan{we conclude that}
PGPRec successfully leverages \zixuan{the} contrastive learning loss to effectively pre-train a graph encoder and further enhances the \zixuan{recommendation} performance by \zixuan{leveraging} the personalised graph prompts \yy{on two of the four datasets used.}
\yy{As a result, our PGPRec framework successfully}
\yy{enriches} the user representations in the target domain, \yy{\yo{by} leveraging the knowledge from the pre-trained model.}


\subsection{Ablation Study (RQ2)}\label{sec:ablation}

\begin{table*}[tb]
\centering
\caption{\yzx{PGPRec performance in terms of Recall@10 and NDCG@10 on the used datasets. 
The best \zixuan{performance} is highlighted in bold and the second best result is highlighted with \zixuan{an} underline.
$^{*}$ denotes a significant difference compared to the result of \yy{PGPRec} using the paired t-test with the Holm-Bonferroni correction for $p<0.05$.}}
\begin{adjustbox}{width=\linewidth}
\begin{tabular}{ccccccccc}
\toprule
\multirow{1}{*}{\textbf{Dataset}} & \multicolumn{2}{c}{Elec-Phone} & \multicolumn{2}{c}{Sport-Cloth} & \multicolumn{2}{c}{Sport-Phone} & \multicolumn{2}{c}{Elec-Cloth}\\ 
\cmidrule(lr){1-1} \cmidrule(lr){2-3} \cmidrule(lr){4-5} \cmidrule(lr){6-7} \cmidrule(lr){8-9}
Methods & Recall@10 & NDCG@10  & Recall@10 & NDCG@10  & Recall@10 & NDCG@10 & Recall@10 & NDCG@10\\
\midrule
PGPRec-BPR & ${0.4851}^{*}$  & ${0.3006}^{*}$ & ${0.5332}^{*}$ & ${0.3149}^{*}$ & ${0.4361}^{*}$  & ${0.2593}^{*}$ & ${0.3918}^{*}$ & ${0.2603}^{*}$\\
\midrule
PGPRec-soft & ${0.4716}^{*}$  & ${0.2845}^{*}$ & ${0.5216}^{*}$ & ${0.3063}^{*}$ & ${0.4468}^{*}$  & ${0.2610}^{*}$ & ${0.4033}^{*}$ & ${0.2662}^{*}$\\
\midrule
PGPRec-hard & ${0.5172}^{*}$  & ${0.3244}^{*}$ & ${0.5532}^{*}$ & ${0.3193}^{*}$ & ${0.4537}^{*}$  & ${0.2856}^{*}$ & ${0.4231}$ & ${0.2814}^{*}$\\
\midrule
PGPRec & \textbf{0.5213}  & \textbf{0.3386} & \textbf{0.5678} & \textbf{0.3281} & \textbf{0.4607}  & \textbf{0.3086} & \textbf{0.4233} & \textbf{0.2854}\\
\bottomrule
\label{tab:ablation}
\end{tabular}
\end{adjustbox}
\end{table*}

\yzx{To investigate the impact of each component of our PGPRec framework, \yo{in Table~\ref{tab:ablation}},
\yy{we compare the results of PGPRec to its variants that employ different types of personalised prompts (soft/hard) as well as PGPRec variants with varying pre-training loss functions (BPR/contrastive loss). \yo{In particular, we} aim to examine the impact of these components on the overall performance of the framework. Table~\ref{tab:ablation} presents the \yo{effect} of these components on the overall performance of the framework.} 
In particular, we evaluate the statistical significance of \yo{the difference in performance}
 between PGPRec and its variants with the paired t-test ($p$ < 0.05).
We first compare PGPRec to its variants with different types of personalised prompts (PGPRec-soft and PGPRec-hard), which sorely use soft and hard prompts, respectively.
From the table, we observe that for all four used datasets, PGPRec significantly outperforms PGPRec-soft and PGPRec-hard in 94\% of the cases
\yo{(15 out of 16 instances).}
This observation indicates the effectiveness of combining both hard and soft prompts for better user/item \yo{representations} by transferring the knowledge from a source domain and tailoring the user/item \yo{representations} to a target domain. \yo{On the other hand, comparing the variants} PGPRec-soft and PGPRec-hard, 
we observe a noticeable performance gap between the two, with PGPRec-hard outperforming PGPRec-soft \yy{on all datasets and metrics}.
This finding indicates that the neighbouring-items carry more informative content than \yo{the} randomly initialised \yz{learned} vectors.
\yy{Hence, these hard prompts}
can better assist user \yy{representations} in adapting
to the target domain. This \yy{comparison between PGPRec-soft and PGPRec-hard} 
highlights the importance of leveraging the appropriate item-level information, \yy{as provided by hard prompts,} to enhance \yo{the} recommendation performance in cross-domain settings.
\yo{Furthermore}, we also compare PGPRec to PGPRec-BPR, which uses \yo{the} BPR loss as \yo{a} training loss \yy{function} during the pre-training stage while PGPRec employs \yo{a} contrastive loss when pre-training the graph encoder.
From Table~\ref{tab:ablation}, we observe that PGPRec \yy{significantly} outperforms PGPRec-BPR by a large margin across all used datasets, 
demonstrating the rationality of incorporating contrastive learning to ensure \zx{an effective} pre-training. 
\yo{The observed enhanced performance highlights} the \yy{importance} of leveraging contrastive learning to extract valuable knowledge from the source domain, \yo{and} ultimately benefiting the downstream target domain.}

\begin{figure}[tb]
    \begin{subfigure}[t]{1\linewidth}
        \hspace{1.6cm}
        \includegraphics[width=0.7\linewidth]{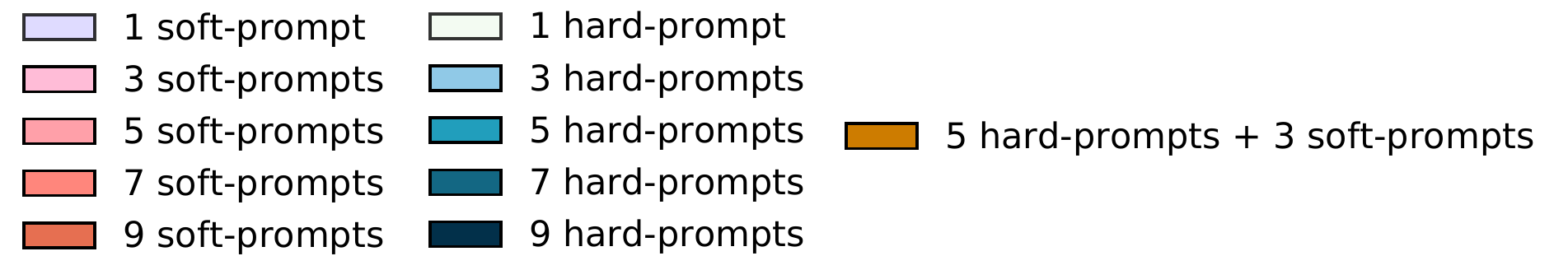}
    \end{subfigure}
    
    \begin{subfigure}[t]{.49\linewidth}
        \includegraphics[clip,trim={0.3cm 0cm 3.8cm 3cm},width=1\linewidth]{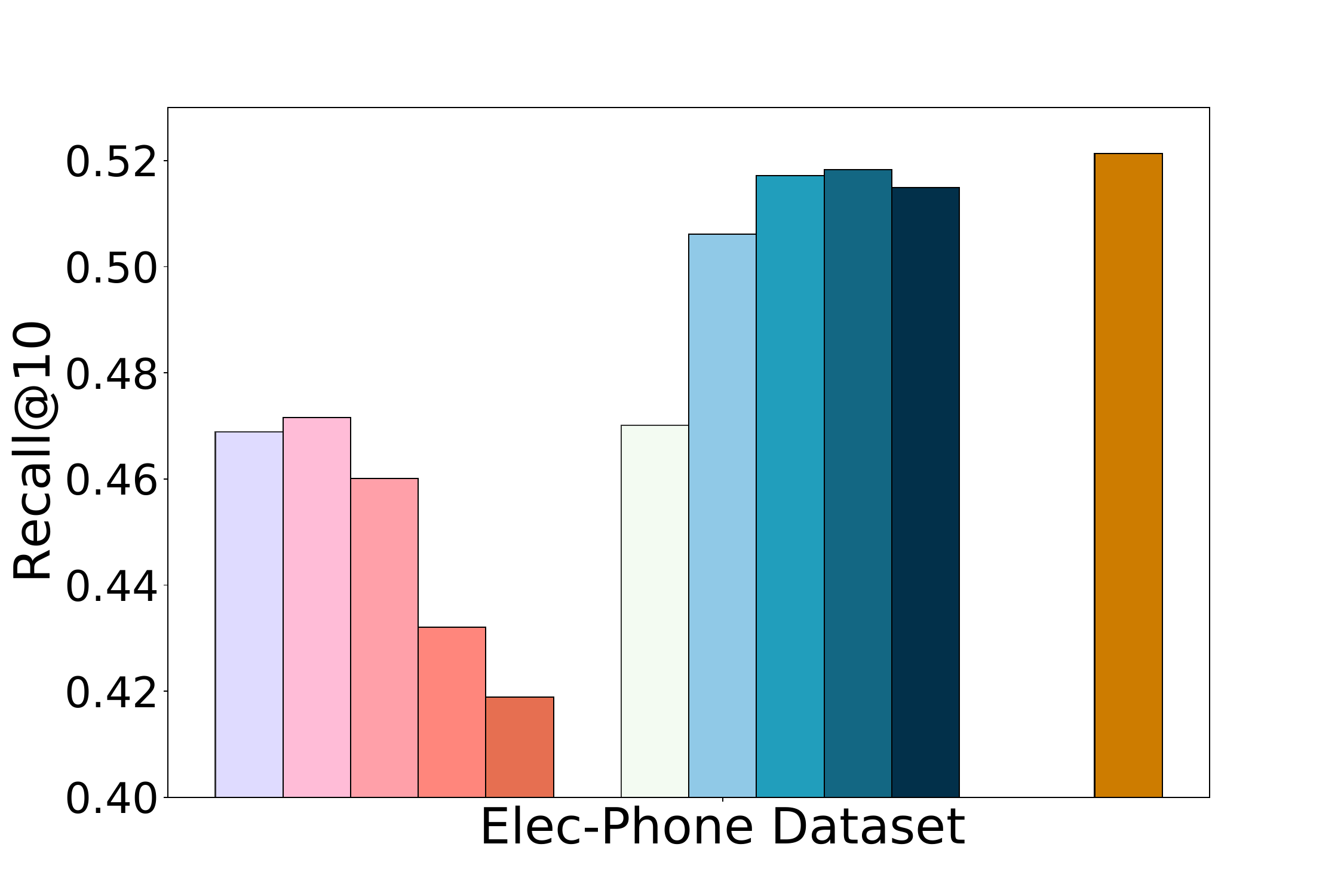}
    \end{subfigure}
    \begin{subfigure}[t]{.49\linewidth}
    \centering
    \includegraphics[trim={0.3cm 0cm 3.8cm 3cm},clip,width=1\linewidth]{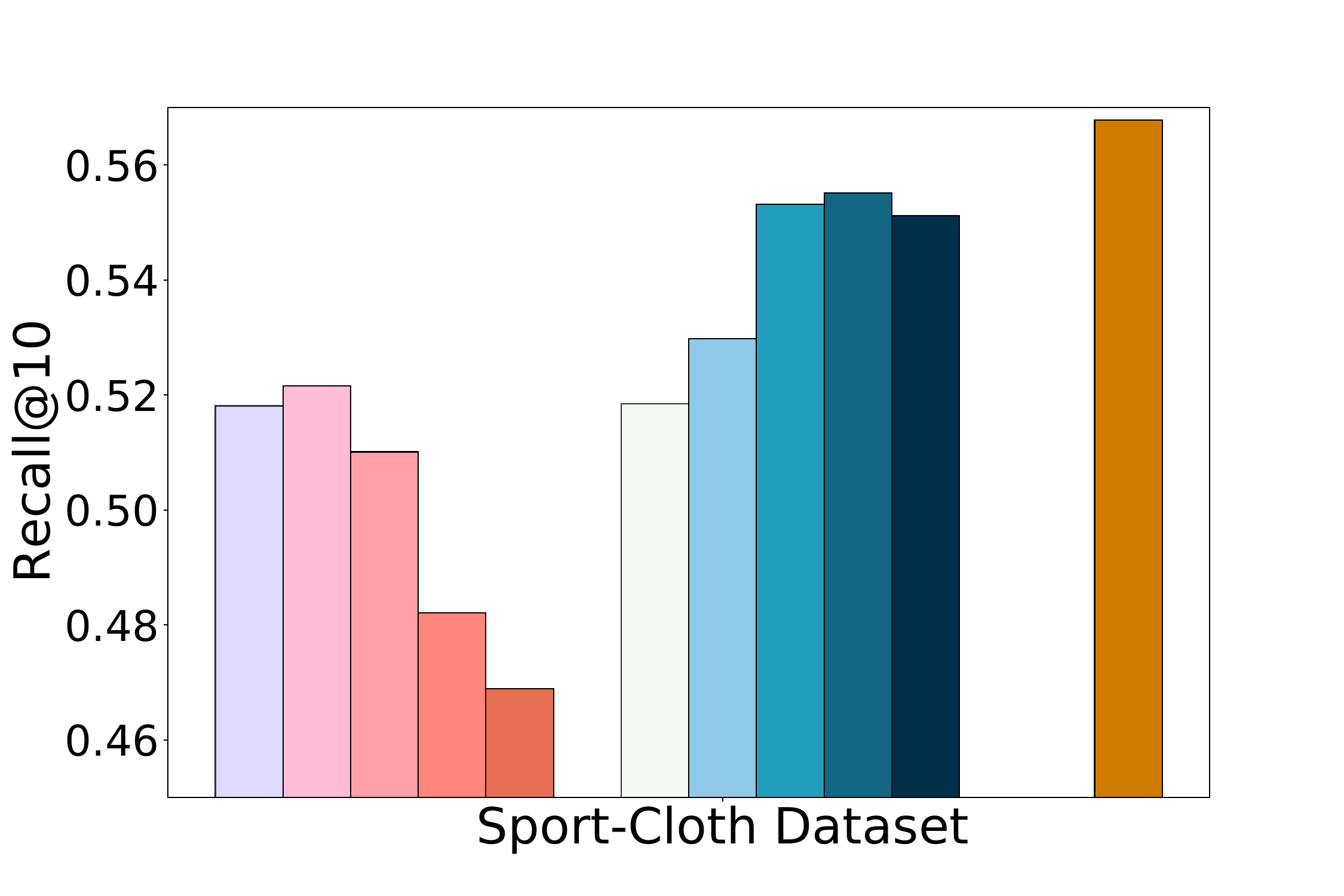}
    \end{subfigure}
    \begin{subfigure}[t]{.49\linewidth}
        \includegraphics[clip,trim={0.3cm 0cm 3.8cm 3cm},width=1\linewidth]{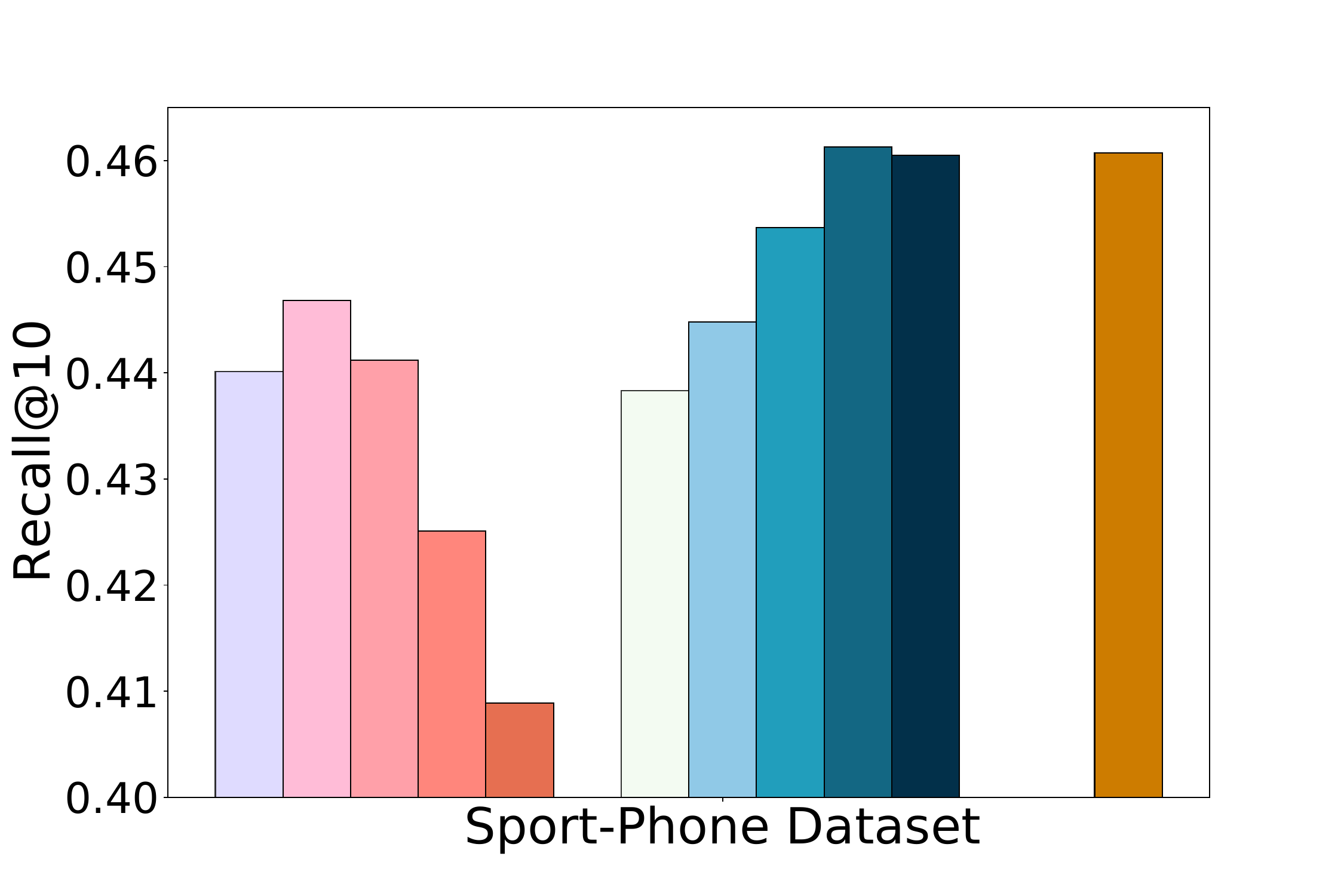}
    \end{subfigure}
    \begin{subfigure}[t]{.49\linewidth}
    \centering
    \includegraphics[trim={0.3cm 0cm 3.8cm 3cm},clip,width=1\linewidth]{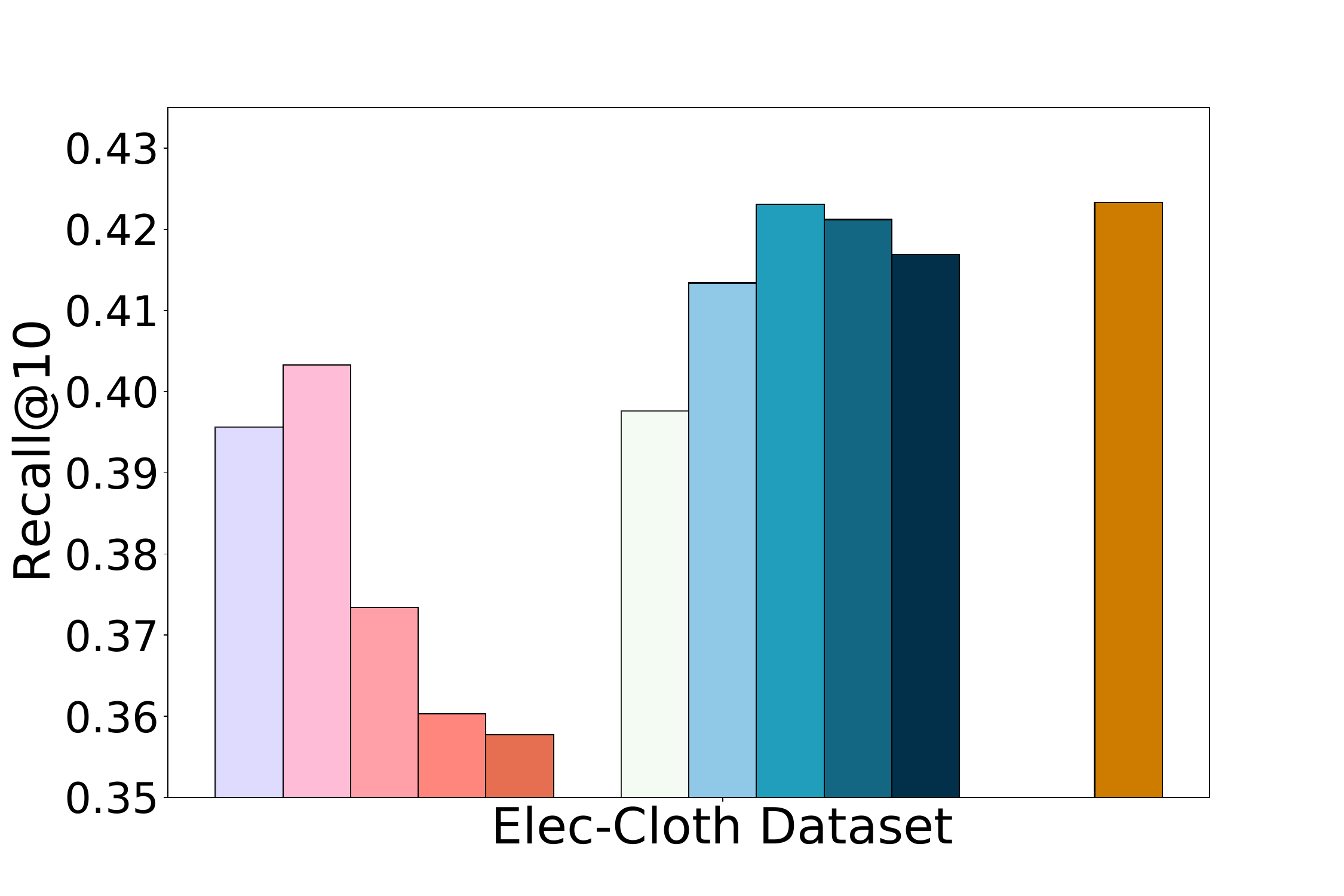}
    \end{subfigure}    
    \centering
    \caption{Performance of \zixuan{the} PGPRec variants in terms of Recall@10 on four Amazon Review datasets.
    } 
\label{fig:ablation}
\end{figure}

To \yzx{further} \zixuan{investigate} the \zixuan{impact} of different combinations of personalised prompts, 
\zixuan{Figure~\ref{fig:ablation} \zixuan{shows how the performance of PGPRec changes as we use different numbers and combinations of \zx{hard and soft} graph prompts}}. 
\zixuan{In particular, Figure~\ref{fig:ablation} shows the results of} \zixuan{a} comprehensive ablation study \zixuan{using} different combinations of hard prompt nodes and soft prompt nodes. 
\zx{For ease of illustration \io{and visual clarity},}
in addition to the performance of PGPRec as we vary the number of used hard and soft prompts, the figure only reports the best-performing combination of soft and hard prompts, which was consistently the same across the used datasets.
\zxy{\io{Note that} we use Recall@10 to report the performance of different PGPRec variants, \io{since using NDCG@10 also leads to the same conclusions} across the used datasets.}
\zixuan{In particular,} Figure~\ref{fig:ablation} shows that the PGPRec variant (the bar in \textit{orange}) \zixuan{that} combines five hard prompt nodes and three soft prompt nodes achieves the best performance \zixuan{in both the Elec-Phone and Elec-Cloth datasets. This promising performance}
is \zixuan{due} to the supplement of \zixuan{the neighbouring-items} from \zixuan{the} \io{available} metadata \io{in the used datasets} \io{as well as} the {use of a} continuous vector as a prompt.
For the PGPRec variants \io{that} only use the hard prompts (bars in \textit{blue}) in Figure~\ref{fig:ablation}, the performance of PGPRec improves with more neighbouring-items, which can be viewed as \zixuan{if} more adjacent nodes of a target node \zixuan{are added/taken into consideration}. However, the improvement becomes marginal \zixuan{when} the number of hard prompt nodes \zixuan{increases to} over five. 
In particular, we can observe that the performance of the variant with seven hard prompts is on \zixuan{a} par with the variant in \textit{orange} on the Sport-Phone and Elec-Cloth \io{datasets}, which combines both hard and soft prompts. 
This further \zixuan{supports} that \zixuan{the} \io{hard} neighbouring-items \zixuan{contribute} most in the personalised graph prompts while the contribution of the soft \io{prompts} is marginal.
\zixuan{By comparing \zixuan{the results observed on the} Elec-Phone and Elec-Cloth datasets, we find that the best number of hard prompts \zixuan{depends on the scale} of the target domain dataset, which \zixuan{indicates} that a larger scale \zixuan{target} dataset always \zixuan{requires} more prompts to facilitate more divergent user/item representations.}
Moreover, \zixuan{we observe from Figure~\ref{fig:ablation} that} the performance of CDR is further enhanced by the \zixuan{addition} of soft prompt nodes. 
One possible reason is that embedding soft prompt nodes \io{provides} \zixuan{an} additional supervised signal \zixuan{when} optimising \zixuan{through} contrastive learning. \zixuan{Indeed, } through contrastive learning, the collaborative signal is further enhanced by mining graph data in a self-supervised manner.
However, \zixuan{as observed in Figure~\ref{fig:ablation}}, a large number of soft prompt nodes impedes the CDR recommendation \zixuan{performance}, which {shows} the difficulty for the graph model to train a useful user/item embedding from scratch.

Hence, \zixuan{in}  answer \zixuan{to} RQ2, \zixuan{we conclude that} PGPRec successfully leverages hard graph prompts to learn effective user/item representations. \zixuan{It also} further enhances performance by optimising the \yz{learned} soft prompt nodes' \zx{embeddings}.


\begin{figure}[tb]
    \begin{subfigure}[t]{0.49\linewidth}
        \includegraphics[width=1\linewidth]{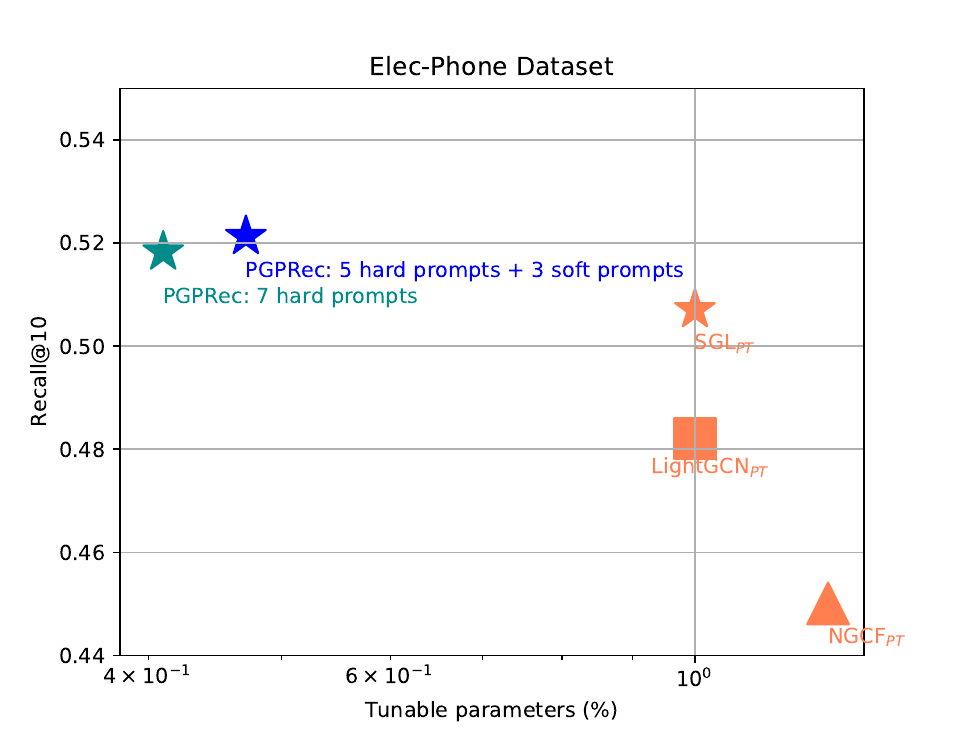}
    \end{subfigure}
    \begin{subfigure}[t]{0.49\linewidth}
        \includegraphics[width=1\linewidth]{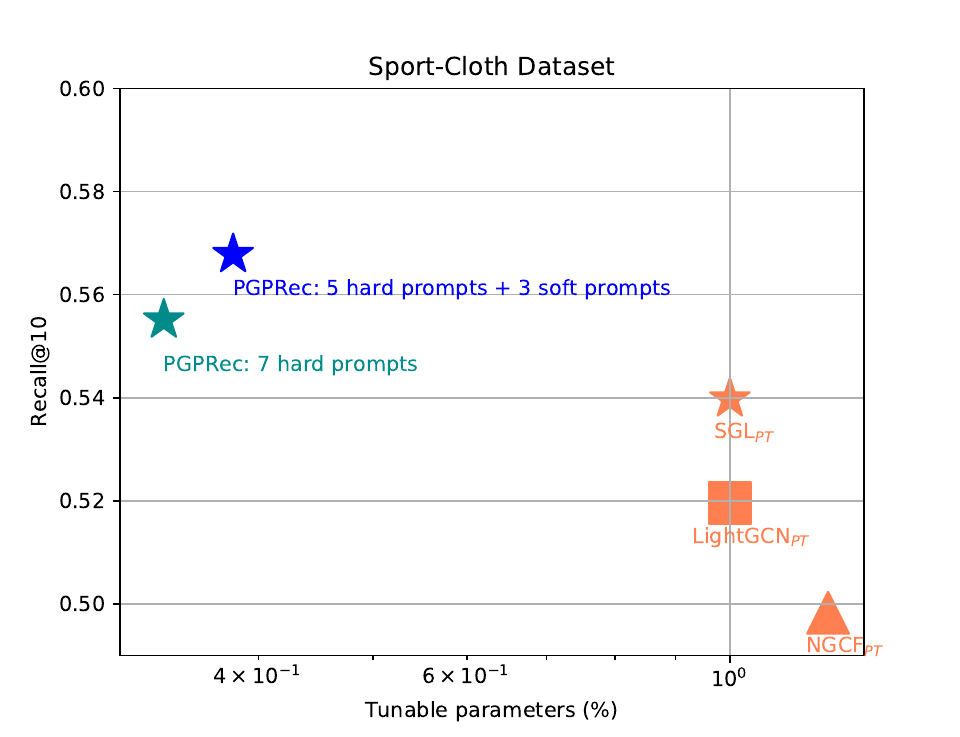}
    \end{subfigure}
        \begin{subfigure}[t]{0.49\linewidth}
        \includegraphics[width=1\linewidth]{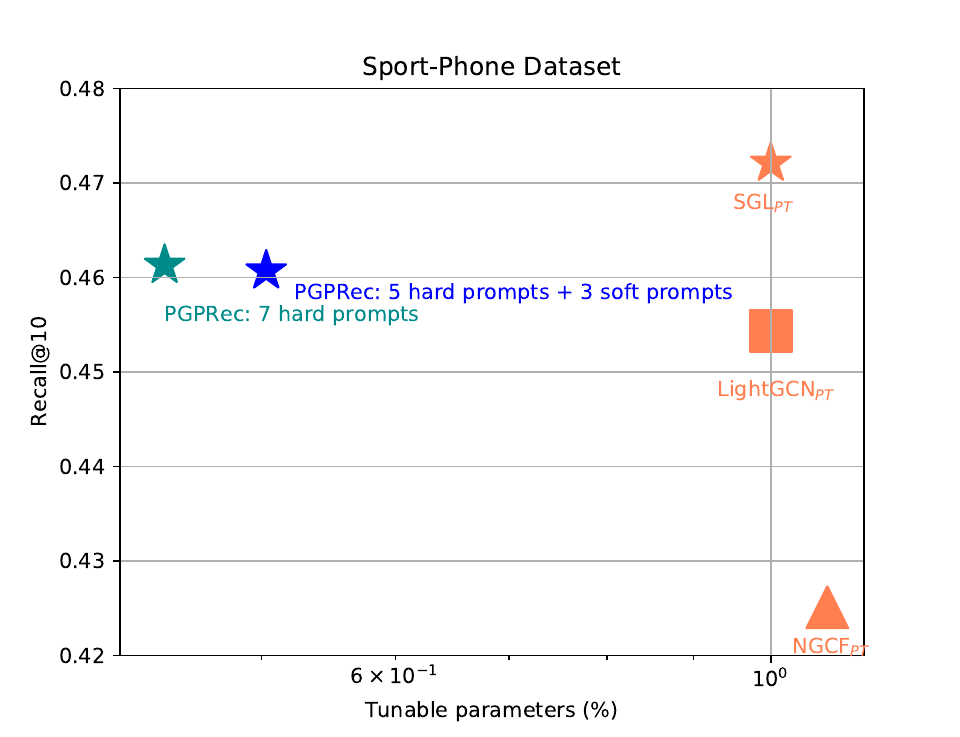}
    \end{subfigure}
    \begin{subfigure}[t]{0.49\linewidth}
        \includegraphics[width=1\linewidth]{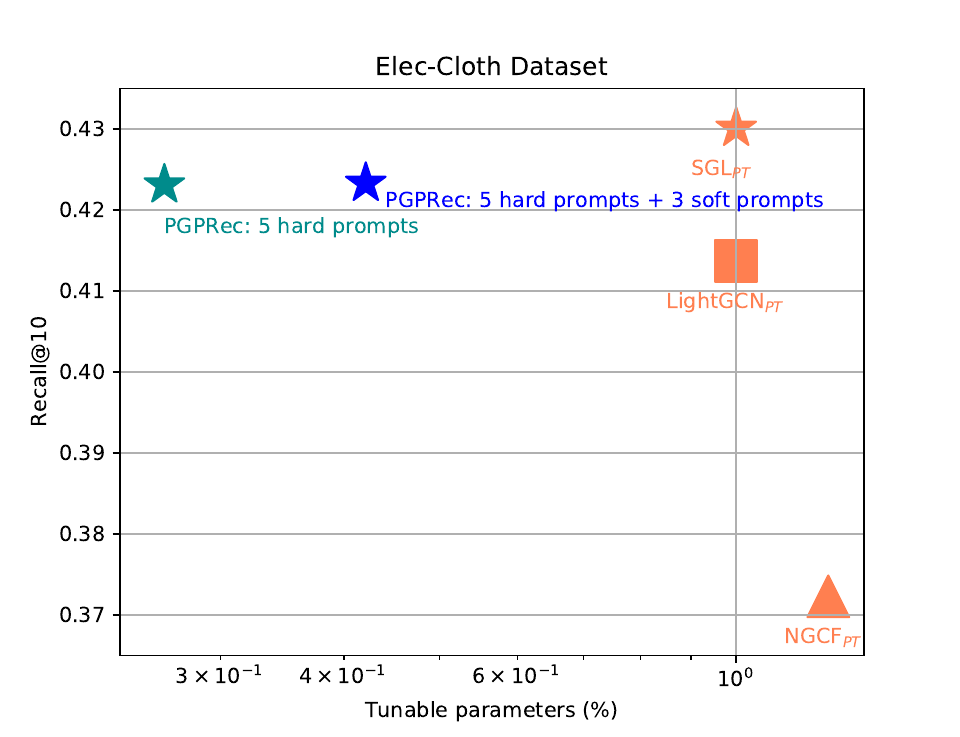}
    \end{subfigure}  
    \caption{\ziy{Performance of different tuning methods on the four used \io{datasets} with the pre-trained GNN models}}
\label{fig:ee}
\end{figure}


\subsection{Efficiency of Graph Prompt-tuning (RQ3)}
\zx{To answer RQ3, we investigate the superiority of personalised graph prompts in terms of parameter efficiency (Section~\ref{sec:pe}) and tuning efficiency (Section~\ref{sec:te}).}
\subsubsection{Parameter Efficiency}\label{sec:pe}
\zixuan{\io{First}, we investigate how many tuned parameters can be reduced 
by using \io{our} personalised graph prompt-tuning \zxy{on all four used datasets.} 
\zxx{Specifically, we examine the \io{parameter efficiency} by comparing the number of \zxx{tuned} parameters of \io{the} personalised graph prompts \io{with} the number of fine-tuned \io{parameters} of \io{the} GNN baselines.}}
\zx{In particular, we use the two most effective PGPRec variants \io{as} reported in Section~\ref{sec:ablation} and calculate the number \zxy{of} \zxx{tuned} parameters.}
\zx{Since efficiency and effectiveness are both critical in our study}, we first compare the calculated number of \zxx{tuned} parameters \io{in} PGPRec \io{as well as in} \io{the} \io{conventionally} fine-tuned GNN baselines \zxy{in Section~\ref{sec:baseline}} \zx{(NGCF$_{PT}$, LightGCN$_{PT}$ and SGL$_{PT}$)\footnote{Recall, that in cross-domain recommendation,  all the pre-trained parameters of the GNN recommenders are fine-tuned in the target domain, while in PGPRec, we only fine-tune the parameters from the personalised graph prompts in the target domain. }.}
\zx{Figure~\ref{fig:ee} shows \io{a comparison of} the effectiveness and efficiency \io{of}  \io{the fine-tuned} GNN \io{baseline} models \io{in comparison to} the variants of PGPRec}.
\zx{For conciseness, we use Recall@10 to report the effectiveness of PGPRec and \io{the} used baselines in Figure~\ref{fig:ee}, \io{since the use of NDCG@10 also leads to the same conclusions} across the used datasets}. 
\zxy{As shown in Figure~\ref{fig:ee}, LightGCN$_{PT}$ and SGL$_{PT}$ have the same number of tuned parameters but less so than NGCF$_{PT}$, which has \io{additional} \yz{learned} weight matrices compared to LightGCN$_{PT}$ and SGL$_{PT}$ (see Section~\ref{sec:baseline}). \io{Note also that SGL$_{PT}$ has the best effectiveness among the baselines}.}
\zxy{We observe that \zxy{the two} PGPRec variants
need only 41\% and
47\% of the required fine-tuned parameters of SGL$_{PT}$ to achieve \zxy{an even \io{higher}} effectiveness \zxy{on the Elec-Phone dataset}}. 
\zx{Similarly}, \zxy{the two} PGPRec variants only \zxy{need} 33\% and 38\% \zx{of the tuned parameters \zxy{required by} SGL$_{PT}$ on the Sport-Cloth dataset.}
\zxy{Next, we also compare our PGPRec variants with NGCF$_{PT}$. 
\zxy{Consider the Elec-Phone dataset as an example}. Figure~\ref{fig:ee} shows that the two PGPRec variants \zxy{need} only 34\% and 39\% of \io{the} required fine-tuned parameters of NGCF$_{PT}$, respectively.}
\zxy{Similarly on the Sport-Cloth dataset, the two PGPRec variants need 27\% and 31\% of the fine-tuned parameter required by NGCF$_{PT}$, respectively.}
\zxx{Similar conclusions \io{can also be observed} on the Sport-Phone and Elec-Cloth datasets, \io{demonstrating the parameter efficiency of PGPRec}.}
\io{Overall, the results in terms of both} \zx{ effectiveness and efficiency on both datasets} \zx{demonstrate} that our personalised graph prompt-tuning is \zx{more parameter-efficient \io{than}} \zx{the conventional} fine-tuning \io{when} transferring pre-trained knowledge to a target domain. 
\zxy{Meanwhile, our PGPRec framework can achieve a competitive performance compared to the fine-tuned GNN baselines on both datasets.}

\begin{figure}[tb]
    \begin{subfigure}[t]{0.49\linewidth}
        \includegraphics[width=1\linewidth]{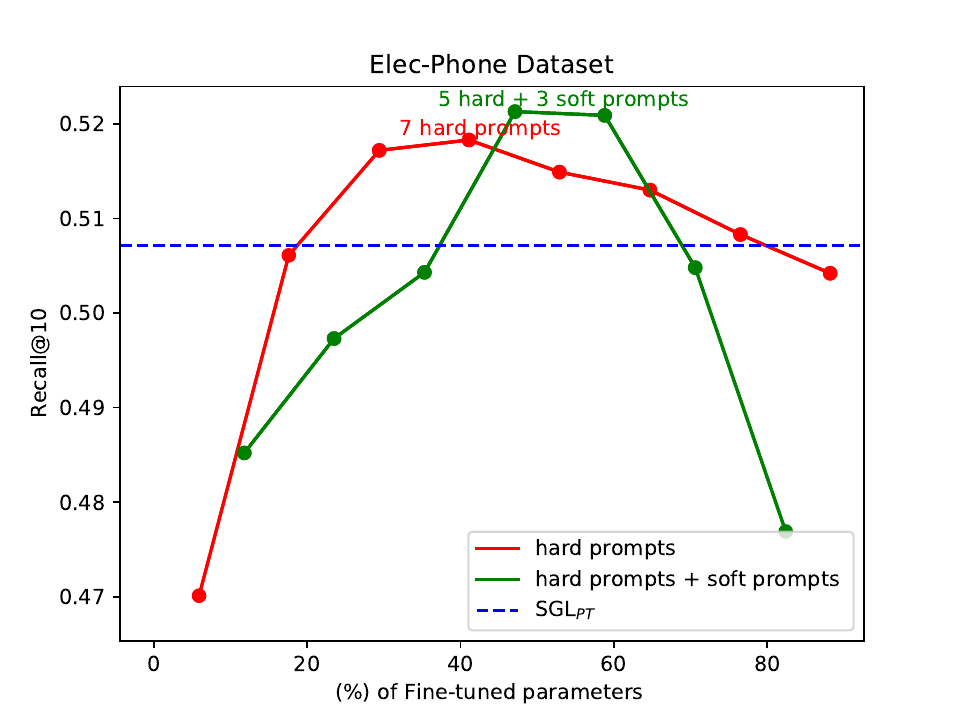}
    \end{subfigure}
    \begin{subfigure}[t]{0.49\linewidth}
        \includegraphics[width=1\linewidth]{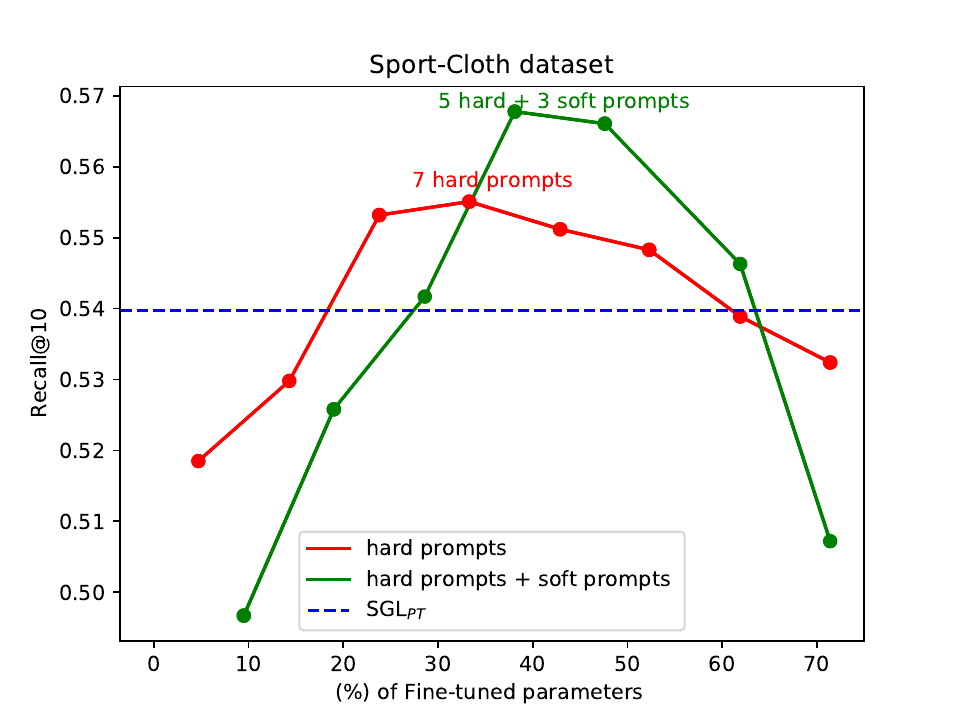}
    \end{subfigure}
    \begin{subfigure}[t]{0.49\linewidth}
        \includegraphics[width=1\linewidth]{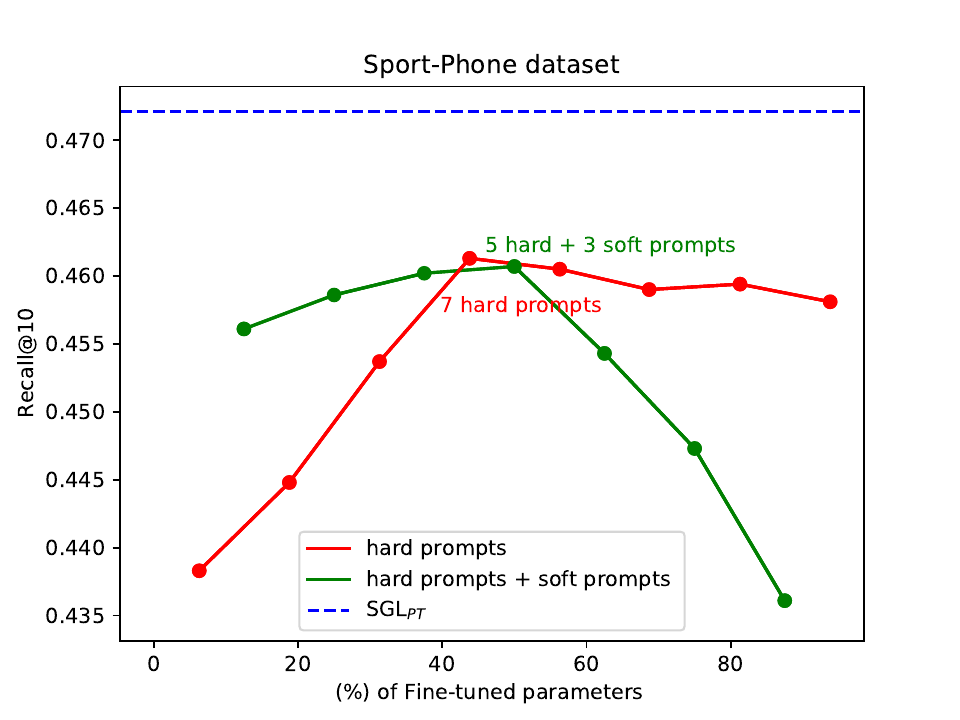}
    \end{subfigure}
    \begin{subfigure}[t]{0.49\linewidth}
        \includegraphics[width=1\linewidth]{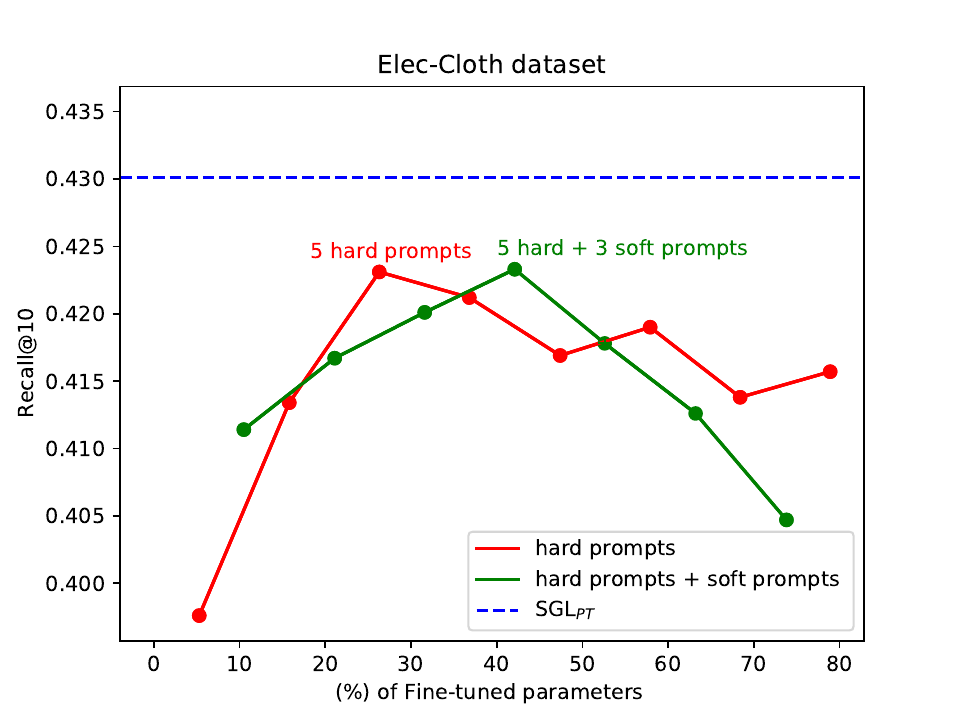}
    \end{subfigure}    
    \caption{
    \yz{A parameter comparison between two types of PGPRec variants: one with only hard prompts and another that combines hard and soft prompts. The peak performances of the two types of PGPRec variants, denoted by the red and green lines respectively, are highlighted with text annotations.}
    \yz{A point marker, represented as a red dot for the PGPRec variant with hard prompts and a green dot for the PGPRec variant that combines hard and soft prompts, indicates that the corresponding performance is significantly equivalent over the SGL$_{PT}$ baseline using the TOST test.}}
\label{fig:cdr_eff}
\end{figure}

\final{To investigate the parameter-efficiency impact of both hard and soft graph prompts within our PGPRec framework, we conduct a comparative analysis on the number of tuned parameters, taking into account the different types of graph prompts used.}
\yz{This comparative analysis encompasses two types of PGPRec variants - one exclusively employs hard prompts (indicated by the red line) while the other uses a combination of hard and soft prompts (indicated by the green line) - as well as the fine-tuned SGL$_{PT}$ baseline.}
\zxy{Note that we use SGL$_{PT}$ for comparison in this experiment because it is the most effective baseline.}
\zxy{For a fair comparison}, both the PGPRec variants and SGL$_{PT}$ use the same GNN encoder (i.e LightGCN).
\zx{Figure~\ref{fig:cdr_eff} shows the results of \io{the two} PGPRec \io{variants} \io{in comparison to the} fine-tuned SGL$_{PT}$ \io{baseline} on all four used datasets.}
\zx{From Figure~\ref{fig:cdr_eff}, we observe that the \zxx{effectiveness} 
of \io{the} \zxy{variant that uses seven hard prompts} \zxy{(in red)}
is on a par with the variant \io{that} combines five hard prompts with three soft prompts \zxy{(in green)} while reducing the \zxy{number of} tuned parameters on \zxy{the x-axes of} all four \io{used} datasets.
This result \io{suggests} that the hard prompts 
\zxy{are overall more efficient in estimating} the users’ preferences in the target domain.}
\zx{\io{In particular},} the PGPRec variant \io{that} \zixuan{leverages} seven hard prompts only needs 41\% and 33\% of the tuned parameters of SGL$_{PT}$ to achieve an even better performance \io{on both the Elec-Phone dataset and \zxy{the} Sport-Cloth dataset, respectively}. This means \io{that the} personalised prompt-tuning \io{allows to} reduce 59\% and 67\% of \io{the} parameters \io{in comparison to using SGL$_{PT}$}, \zxy{while \io{also} attaining an improved effectiveness}.
\zxx{Similarly, the PGPRec variant that uses seven hard prompts can reduce by 54\% the number of \zxx{tuned} parameters of SGL$_{PT}$ on \io{the Sport-Phone dataset}, \io{and} the PGPRec variant that uses five hard prompts can reduce by 74\% the number of tuned parameters of SGL$_{PT}$ on the Elec-Cloth dataset, while ensuring a competitive performance on both datasets.} 
\zx{\io{To verify that the reduction in the number of \zxx{tuned} parameters does not lead to a significant degradation of effectiveness in comparison to SGL$_{PT}$ on all 4 used datasets,} we apply a two one-sided equivalence test (TOST). This test is performed to ascertain \io{an} effectiveness equivalence with the SGL$_{PT}$ model.}
\zxy{In Figure~\ref{fig:cdr_eff}, \zxx{a} point marker \zxx{(corresponding to a green or red dot)} indicates that the corresponding performance is significantly equivalent with SGL$_{PT}$ using the TOST test.}
The \io{TOST} results 
validate our \io{argument} that only tuning on the \zxx{parameters of graph} prompts with all \zixuan{other} pre-trained parameters \io{remaining} unchanged can achieve \zixuan{a competitive \zixuan{performance} \zxy{in comparison to SGL$_{PT}$}} \zxy{on all used datasets}.

\subsubsection{Efficiency of Tuning Time}\label{sec:te}
\zx{We now analyse the time efficiency of \io{PGPRec}. 
As described in Section~\ref{sec:implement}, \io{for a fair comparison}, we use one Geforce RTX 2080Ti GPU to conduct the \io{time efficiency} experiments. 
For conciseness, we report the tuning \io{times} \io{for} PGPRec and \zxy{the two most effective baselines} (LightGCN$_{PT}$ and SGL$_{PT}$) on the Elec-Phone dataset, \io{since the conclusions on the other datasets are similar}.}
\zixuan{Table~\ref{tab:cdr_eff} \zixuan{compares the efficiency of \io{various}} PGPRec variants on the Elec-Phone dataset in terms of tuning time. Specifically, we \zixuan{compare} the \zixuan{effective} PGPRec variants \zixuan{from Section~\ref{sec:ablation}}, namely the variants with seven hard prompts \io{(denoted by PGPRec$_{hard}$)}, three soft prompts \io{(denoted by PGPRec$_{soft}$) and \zixuan{the} mixture of five hard prompts along with three soft prompts} \io{(denoted by PGPRec$_{mixture}$)}  \zixuan{to} \zxy{LightGCN$_{PT}$ and SGL$_{PT}$.}
\zxy{Note that we do not compare with NGCF$_{PT}$ as it has additional learned weight matrices compared to LightGCN$_{PT}$ and SGL$_{PT}$.}
From Table~\ref{tab:cdr_eff}, we observe that all PGPRec variants have a faster training speed (seconds per epoch) than \zixuan{the} LighGCN$_{PT}$ and SGL$_{PT}$ baselines, which \zixuan{indicates} 
the efficiency of using prompts to tune on a small portion of parameters.
Moreover, PGPRec$_{hard}$ takes the least training time while achieving a competitive performance compared with SGL$_{PT}$ and PGPRec$_{mixture}$. \zixuan{This} demonstrates the efficiency of prompt-tuning \zixuan{compared with} conventional fine-tuning \io{as well as} the effectiveness of hard prompts in enhancing the pre-trained embeddings.
However, Table~\ref{tab:cdr_eff} \io{also} shows that PGPRec$_{soft}$ and PGPRec$_{mixture}$ take more epochs to converge, which indicates the difficulty \zxx{of the graph encoder (LightGCN) in PGPRec} to train useful \zxx{soft graph prompts} from scratch.}

{To answer RQ3, \zxx{according to Figure~\ref{fig:ee} and Figure~\ref{fig:cdr_eff}}, we conclude that PGPRec empowered by personalised graph prompts is more parameter-efficient than fine-tuned GNNs such as NGCF$_{PT}$, LightGCN$_{PT}$ and SGL$_{PT}$. 
\zx{Moreover, \io{as shown in} Table~\ref{tab:cdr_eff}, we \final{find} that only using hard graph prompts helps to further improve the efficiency of tuning time while \io{maintaining} \zxy{a \io{competitive} effectiveness}.}}

\begin{table}[tb]
\centering
\small
\caption{Efficiency comparison on the Elec-Phone Dataset in \io{terms of} tuning time. PGPRec$_{hard}$, PGPRec$_{soft}$ and PGPRec$_{mixture}$ refer to the PGPRec variants \io{with} hard prompts, soft prompts and mixture prompts, respectively.}
\begin{adjustbox}{width=0.7\linewidth}
\begin{tabular}{l|c c c c}
     \hline
     Model & Time/Epoch & Nbr of Epochs & Training Time & Recall@10 \\
     \hline
     LightGCN$_{PT}$ & 63s & 294 & 309m & 0.4821 \\
     SGL$_{PT}$ & 68s & 138 & 157m & 0.5071\\
     \hline
     PGPRec$_{mixture}$ & 57s & 423 & 402m & 0.5213 \\
     PGPRec$_{soft}$ & 26s & 367 & 159m & 0.4716 \\
     PGPRec$_{hard}$  & 44s &  147 & 108m & 0.5172\\
     \hline
\end{tabular}
\label{tab:cdr_eff}
\end{adjustbox}
\end{table}

\subsection{Cold-start Analysis (RQ4)}

\begin{table*}[tb]
\caption{Cold-start analysis results \zixuan{for} PGPRec and \zixuan{the} SGL$_{PT}$ \zxy{baseline}, \zxy{where} PT is the \io{abbreviation} \zixuan{for} Pre-Training. \zxy{'Cold-start' denotes the cold-start users and 'Regular' denotes the regular users in the used datasets. }
$^{*}$ denotes a significant difference \zxx{between PGPRec and SGL$_{PT}$} using the \zxx{paired t-test with p<0.05}.}
\begin{adjustbox}{width=0.7\linewidth}
    \begin{tabular}{l|c|c|c|c}
      \hline
      \textbf{Datasets} &  &\text{PGPRec (Recall@10)} & \text{SGL$_{PT}$ (Recall@10)} & \text{\%Improv.}\\
      \hline
      \multirow{2}{*}{Elec-Phone}
      & Cold-start & 0.4671*  & 0.4314 & 8.27\%\\
      & Regular & 0.7501* & 0.7343 & 2.15\%\\
      \hline
      \multirow{2}{*}{Sport-Cloth}
      & Cold-start & 0.5185* & 0.4653 & 11.41\%\\
      & Regular & 0.6735* & 0.6564 & 2.60\%\\%
      \hline
      \multirow{2}{*}{Sport-Phone} 
      & Cold-start & 0.4224* & 0.4042 & 4.5\%\\
      & Regular & 0.5483 & 0.5687 & -3.59\%\\%
      \hline
      \multirow{2}{*}{Elec-Cloth}
      & Cold-start & 0.3876* & 0.3764 & 2.98\%\\
      & Regular & 0.6192 & 0.6354 & -2.62\%\\%
      \hline
    \end{tabular}
\end{adjustbox}
\label{tab:cold}
\end{table*}

\ziy{To investigate the effectiveness of transferring knowledge to the cold-start users \io{using our} personalised graph prompts, we examine our PGPRec \zx{framework} in a cold-start scenario.
Specifically, we use the \zxx{most effective} PGPRec variant \zx{from Section~\ref{sec:ablation}}, \io{namely the PGPRec variant with \zxx{five hard prompts and three soft prompts}},
and perform a cold-start analysis on four Amazon Review datasets.
Table~\ref{tab:cold} shows the performances of PGPRec for \io{both the} cold-start and regular users, in comparison to the best \io{used} baseline, SGL$_{PT}$, in terms of Recall@10 
\zxx{(\io{similar trends can be observed with NDCG@10)}.}
The results show that PGPRec benefits the cold-start users more than SGL$_{PT}$ \io{and} significantly \zxy{according to the paired t-test} 
on all \zxy{four} datasets. \io{This observation suggests} that PGPRec successfully leverages the neighbouring-items and \io{the} \yz{learned} vectors as graph prompts to bring useful information to estimate \zxy{a user's preferences.}
\io{In particular}, PGPRec outperforms SGL$_{PT}$ \io{by}
8.27\% and 11.41\% on \io{the} Elec-Phone dataset and \io{the} Sport-Cloth datasets, \io{respectively}, \io{thereby demonstrating} the successful feature transfer from the source domain and \io{the added-value} of personalised graph prompts as additional information. \io{Significant improvements on the cold-start users in comparison to SGL$_{PT}$ are also obtained on \io{the} Sport-Phone and Elec-Cloth datasets, although their scale is smaller (4.5\% and 2.98\%, respectively}.
This \io{latter} observation indicates that the transferred features from the source domain are not \io{sufficiently} informative to enhance the prediction of cold-start users' preference in a distant target domain.}
\zx{\io{In fact}, \io{given the scale of improvements in} \zxy{Table~\ref{tab:cold}}, \io{we observe} that our PGPRec framework \io{actually} benefits the cold-start users more than the regular users}. 
\ziy{For example,  \io{on the Sport-Cloth dataset}, PGPRec improves the performance by 11.41\% for the cold-start users  \io{in comparison to SGL$_{PT}$}, \io{while it only improves the performance by just 2.60\% for the regular users}.} \io{This suggests that PGPRec is \zxx{particularly} helpful in cold-start scenarios}. 
\io{This result suggests that the} PGPRec \zxx{framework} successfully leverages the personalised graph prompts to act as additional adjacent items to a cold-start user, \io{thereby enriching the representations of users with} sparse interactions \final{in the target domain}.

\zx{Overall, in response} to RQ4,  we conclude that \zxx{our} PGPRec \zxx{framework} successfully leverages \zxx{the} pre-trained features and \zxx{the} personalised graph prompts as auxiliary information to \zx{effectively enhance the recommendation performance on cold-start users}.

\section{Conclusions}
In this work, \zixuan{we proposed a Personalised Graph-based Prompt Recommendation (PGPRec) framework to enhance the efficiency of conventional fine-tuning in cross-domain recommendation. 
Specifically, we devised \io{novel} personalised graph prompts to
further \io{enrich} the pre-trained embeddings in the target domain.
In particular, we used the neighbouring-items as hard prompts and \yz{used} randomly initialised embedding vectors as soft prompts. 
\zixuan{We} leveraged the personalised graph prompts to enhance the efficiency of the tuning stage. 
Our results on four cross-domain datasets showed that PGPRec efficiently leverages the prompt-tuning
\zixuan{along with the} \yz{state-of-the-art} graph recommenders \io{and \yz{achieves}} a competitive performance \zixuan{compared} with the strongest baseline (SGL$_{PT}$). 
Moreover, we conducted \zixuan{an} ablation study \zixuan{investigating} different combinations of the personalised graph prompts.
We showed that the hard prompts make a \yz{key and marked} contribution in the tuning stage \yz{while} the soft prompts can provide limited improvement to top-$k$ recommendation in the target domain. Furthermore, we showed that \yz{a} PGPRec \io{variant} with \io{five} hard prompts can \yz{markedly} reduce the \zixuan{number} of the \io{required} \zxx{tuned} parameters \yz{by a large margin} (e.g., a 74\% \io{reduction in the number of tunable parameters of \yz{the strongest baseline} SGL$_{PT}$} \zixuan{on} the Elec-Cloth dataset). By comparing the tuning time \zixuan{\yz{of} the \yz{existing}} fine-tuned GNNs \zixuan{\yz{with} that of} \io{several PGPRec variants}, we showed that \zixuan{our PGPRec framework} tunes faster (e.g., 31\% faster on the Elec-Phone dataset), \zixuan{\zxx{when using only} hard prompts, leading to \yz{a} marked training efficiency}}.
\ziy{\zx{Finally}, we conducted a cold-start analysis to investigate \io{whether} \yz{our proposed} PGPRec \yz{framework} \io{benefits the} cold-start users. The results \io{obtained} on four datasets showed that PGPRec successfully alleviates the cold-start problem, \yz{and achieves} effective cross-domain recommendations (e.g. an 11.41\% \io{performance} improvement on the Sport-Cloth dataset \yz{in comparison to the strongest baseline SGL$_{PT}$}) for cold-start users.}

 

\balance
\bibliographystyle{ACM-Reference-Format}
\bibliography{reference}

\end{document}